\documentclass{article}

\usepackage{arxiv}

\usepackage[utf8]{inputenc} 
\usepackage[T1]{fontenc}    
\usepackage{hyperref}       
\usepackage{url}            
\usepackage{booktabs}       
\usepackage{amsfonts}       
\usepackage{nicefrac}       
\usepackage{microtype}      
\usepackage{lipsum}
\usepackage{graphicx}
\graphicspath{ {./images/} }

\usepackage{natbib}
\usepackage{bm}
\usepackage{hhline}
\usepackage{subcaption}
\usepackage{comment}
\usepackage{titlesec}
\usepackage{color}
\usepackage[table]{xcolor}
\usepackage{tabularx}
\usepackage{rotating}
\usepackage{algorithm}
\usepackage{algpseudocode}
\usepackage{eso-pic}
\usepackage{bbm}

\usepackage[overload]{empheq}
\definecolor{deeppink}{rgb}{1.0, 0.08, 0.58}
\usepackage{fancyhdr}
\usepackage{tcolorbox}
\usepackage{colortbl}
\usepackage{tabularray}
\usepackage{float}

\title{
Cluster-weighted modeling of lifetime hierarchical data for profiling COVID-19 heart failure patients}

\author{
  Luca Caldera \\
  MOX, Department of Mathematics \\
  Politecnico di Milano \\
  Milan, IT \\
  \texttt{luca.caldera@polimi.it} \\
  \And
  Andrea Cappozzo \\
  Department of Statistical Sciences \\
  Università Cattolica del Sacro Cuore \\
  Milan, IT \\
  \texttt{andrea.cappozzo@unicatt.it} \\
  \And
  Chiara Masci \\
  Department of Economics, Management, and Quantitative Methods \\
  University of Milan \\
  Milan, IT \\
  \texttt{chiara.masci@unimi.it} \\
  \And
  Marco Forlani \\
  ARIA S.p.A. - Divisione Digital Information HUB \\
  Milan, IT \\
  \texttt{marco.forlani@ext.ariaspa.it} \\
  \And
  Barbara Antonelli \\
  ARIA S.p.A. - Divisione Digital Information HUB \\
  Milan, IT \\
  \texttt{barbara.antonelli@ariaspa.it} \\
  \And
  Olivia Leoni \\
  U.O. Osservatorio Epidemiologico, DG Welfare \\
  Regione Lombardia \\
  Milan, IT \\
  \texttt{olivia\_leoni@regione.lombardia.it} \\
  \And
  Anna Maria Paganoni \\
  MOX, Department of Mathematics \\
  Politecnico di Milano \\
  Milan, IT \\
  \texttt{anna.paganoni@polimi.it} \\
  \And
  Francesca Ieva \\
  MOX, Department of Mathematics \\
  Politecnico di Milano \\
  Milan, IT \\
  \texttt{francesca.ieva@polimi.it} \\
}

\begin{document}
\maketitle
\begin{abstract}
This study investigates the heterogeneity in survival times among COVID-19 patients with Heart Failure (HF) hospitalized in the Lombardy region of Italy during the pandemic. To address this, we propose a novel mixture model for right-censored lifetime data that incorporates random effects and allows for local distributions of the explanatory variables. Our approach identifies latent clusters of patients while estimating component-specific covariate effects on survival, taking into account the hierarchical structure induced by the healthcare facility. Specifically, a shared frailty term, unique to each cluster, captures hospital-level variability enabling a twofold decoupling of survival heterogeneity across both clusters and hierarchies. Two EM-based algorithms, namely a Classification EM (CEM) and a Stochastic EM (SEM), are proposed for parameter estimation. The devised methodology effectively uncovers latent patient profiles, evaluates within-cluster hospital effects, and quantifies the impact of respiratory conditions on survival. Our findings provide new information on the complex interplay between the impacts of HF, COVID-19, and healthcare facilities on public health, highlighting the importance of personalized and context-sensitive clinical strategies.
\end{abstract}


\section{Introduction}

The COVID-19 pandemic has placed extraordinary pressure on healthcare systems worldwide, intensifying the challenges associated with managing chronic conditions such as heart failure (HF). Patients with pre-existing cardiovascular diseases, including HF, have shown increased vulnerability and increased mortality when infected with SARS-CoV-2 \citep{rey2020heart, bader2021heart, adeghate2021mechanisms}. To comprehensively analyze the complex interplay between COVID-19 and HF, it is therefore crucial to consider the broader clinical and systemic context. This includes accounting for variability in patient-level clinical characteristics and hospital-related factors within a unified quantitative framework capable of capturing their intricate interactions. Such an integrated perspective is especially important for areas severely affected by the pandemic, such as the Lombardy region in Italy, which experienced one of the highest burdens during the early stages of the outbreak. Accurately assessing survival outcomes and isolating the impact of hospital-specific factors in this context requires modeling approaches capable of handling heterogeneous data sources and stratified patient populations.

To address this need, Cluster Weighted Models (CWMs) offer a flexible framework for jointly modeling covariates and outcomes in the presence of latent subpopulations \citep{Gershenfeld1997a,Ingrassia2012, Ingrassia2014}. Although CWMs have proven effective in various domains involving mixed or structured data \citep{Berta2016, Berta2019, Berta2024, Caldera2025}, existing formulations are not designed to handle time-to-event outcomes, a crucial limitation in the context of survival analysis.

Motivated by the problem of profiling HF patients hospitalized for COVID-19 in the Lombardy region during the pandemic, this paper proposes a novel hierarchical survival model that extends the CWM framework to handle nested time-to-event responses. The devised methodology captures cluster-specific relationships between covariates and survival outcomes through a shared frailty term, making it well suited for heterogeneous time-to-event data with an inherent hierarchical structure, such as hospital of admission in medical applications. More specifically, the frailty model introduced by \cite{vaupel1979impact} accounts for the dispersion arising from the hierarchy incorporating a multiplicative factor known as frailty. The frailty can be modeled parametrically, typically with a Gamma or Lognormal distributions, or through a semiparametric approach.  Detailed introductory reviews on this topic can be found in \cite{Abrahantes2007a, Austin2017} and \cite{Balan2020}. 

By incorporating a parametric frailty term into the CWM specification, we propose a novel methodology that enables the joint identification of latent patient clusters with distinct survival patterns, while also uncovering the impact of the known hierarchical structure on survival within each cluster. From a clinical perspective, the main objectives of this study are the following.

\begin{itemize} \item Identify relevant latent subpopulations of HF patients hospitalized for COVID-19, based on their clinical records.
\item Evaluate how different medical facilities influence the hazard of death for individuals with distinct clinical profiles.
\item Investigate the impact of respiratory pathologies on mortality risk between clusters of patients.
\item Integrate these multiple sources of information to produce customized survival curves that stratify death risk by patient profiles, respiratory conditions, and hospital of admission.
\end{itemize}

The remainder of the paper is organized as follows. In Section~\ref{subsec:prob_sett_and_data_descr}, we present the administrative database of the Lombardy region, highlighting the clinical information of the patient that motivated our study. Section~\ref{sec:methodology} details the proposed methodology, including the model setting and the estimation procedure. Section~\ref{sec:applic_and_results} presents the application of the novel method to the administrative database of the Lombardy region, along with the corresponding results. In Section~\ref{sec:sim_study}, we perform a simulation study to assess the performance of the devised methodology under a controlled scenario. Finally, Section~\ref{sec:discussion} concludes the work with key insights and suggestions for future research.

\section{Data Description}
\label{subsec:prob_sett_and_data_descr}
The data set considered originates from the administrative database for health care of the \href{https://www.regione.lombardia.it/wps/portal/istituzionale/HP/DettaglioServizio/servizi-e-informazioni/Enti-e-Operatori/sistema-welfare/Accreditamento/accesso-db-covid/accesso-db-covid}{Lombardy region}, which is responsible for the comprehensive recording and aggregation of various health services. Specifically, we consider patients diagnosed with heart failure (HF) between January 1, 2018, and December 31, 2020. From this group, we select those who were subsequently hospitalized with COVID-19\footnote{A patient is classified as having HF if their records show hospitalizations or ER visits under DRG code 127 (“Heart Failure and Shock”) per the Lombardy Region’s MS-DRG v40 system. This includes primary or secondary diagnoses of HF (ICD-9-CM: 428.*) or related conditions (e.g. ICD-9-CM: 402.01, 402.11, 402.91). Hospitalizations for COVID-19 are identified using a dedicated flag based on regional coding guidelines during the pandemic.} in the Lombardy region between January 31, 2020 and June 18, 2021. The hierarchy is represented by the hospitals in which patients are admitted. Hospitals with fewer than 50 patients were excluded to ensure sufficient sample sizes for a reliable estimate of frailty-specific parameters. Patients with hospital transfers for medical reasons or repeated infections were also excluded. Our final sample includes $N=3086$ patients hospitalized in $J=32$ different hospitals in the Lombardy region. For each patient, the data set provides details about personal information, the admission facility, and clinical status. For the time-to-event analysis, we selected a 90-day observation window. Patients who survived or died after this period were treated as censored. The patient-level variables we include are:
\begin{itemize}
    \item  \textbf{Time}: it represents either the survival time or the observation period, depending on the patient’s status. 
    \item \textbf{Status}: a binary variable with a value of 1 if the event occurs within the time window and 0 otherwise.
    \item \textbf{Age}: The age of the patient at the time of hospital admission.
    \item \textbf{Gender}: The biological sex of the patient.
    \item \textbf{Modified Multisource-Comorbidity Score (MCS)}: The MCS, introduced by \citet{corrao2017developing}, is a validated index that summarizes comorbidity, defined as the cumulative burden of diseases not related to the primary diagnosis of a patient. It serves as a reliable proxy for overall health status. In this study, we employ a modified version of the MCS, previously used by \citet{Caldera2025}, which excludes respiratory diseases of particular medical relevance to COVID-19 infection. The resulting score is a quantitative variable that theoretically varies from 0, indicating no comorbidities, to 150, corresponding to the presence of all the conditions considered. Details on the computation of this index are provided in Supplement B.
    \item \textbf{Respiratory diseases}: These are represented as dichotomous variables that indicate the presence or absence of specific respiratory conditions. In detail,
    \begin{itemize}
        \item Pneumonia (PNA): An acute inflammation of the lungs caused by infection.
        \item Respiratory Failure (RF): A condition in which blood oxygen levels drop to critical low levels or carbon dioxide levels rise dangerously high.
        \item Chronic obstructive pulmonary disease (COPD): A common lung disease that restricts airflow and causes breathing difficulties. It includes conditions such as emphysema and chronic bronchitis. In COPD, the lungs may be damaged or obstructed by mucus. Common symptoms include coughing (often with phlegm), shortness of breath, wheezing, and fatigue.
        \item Bronchitis (BRH): Inflammation of the bronchi typically due to infection, resulting in irritation and swelling.
    \end{itemize}
\end{itemize}
A brief summary of the most important descriptive statistics is provided in Table~\ref{tab: cont summ} for continuous variables and Table~\ref{tab: cat summ} for categorical variables, respectively.

Disentangling heterogeneity in survival times due to respiratory diseases and hospital effects across latent patient clusters presents a complex learning challenge. We address this by introducing a novel multilevel cluster-weighted model for lifetime data, which is detailed in the following section.

\begin{table*}[htbp]
\caption{Summary of Continuous Variables in Patients with Heart Failure and COVID-19 – Lombardy Region Dataset. Note: Descriptive statistics of variable Time refer only to observations for which the event is observed (i.e., Status = 1).}
\renewcommand{\arraystretch}{1.6}
\centering
\begin{tblr}{c c c c c c c c}
\hline
  \textbf{Variable} & \textbf{Mean} & \textbf{Std.Dev} & \textbf{$\boldsymbol{1}^{\textbf{st}}$ \textbf{Quartile}} & \textbf{Median} & \textbf{$\boldsymbol{3}^{\textbf{rd}}$ \textbf{Quartile}} & \textbf{Min} & \textbf{Max} \\ 
 \hline \hline
 Time [Days] & 19.2 & 18.04 & 6.0 & 12.0 & 26.0 & 1.0 & 89.0  \\
\hline
Age & 81.361 & 8.273 & 76.0 & 82.0 & 88.0 & 60.0 & 106.0  \\
\hline
MCS & 11.468 & 7.768 & 6.0 & 10.0 & 16.0 & 0 & 57.0 \\
\hline
\end{tblr}
\label{tab: cont summ}
\end{table*}

\begin{table*}[htbp]
\caption{Summary of Categorical Variables in Patients with Heart Failure and COVID-19 – Lombardy Region Dataset.}
\renewcommand{\arraystretch}{1.6}
\centering
\begin{tblr}{l l c}
\hline
 \textbf{Variable} & \textbf{Levels} & \textbf{Frequency} \\
\hline \hline
\SetCell[c=1]{l,black!10} Status & Deceased & 44.75\% \\
 & Alive & 55.25\% \\
\hline
 \SetCell[c=1]{l,black!10} Gender & Male & 55.83\% \\
 & Female & 44.17\% \\
\hline
\SetCell[c=1]{l,black!10} COPD & Present & 24.21\% \\
 & Not Present & 75.79\% \\
\hline
\SetCell[c=1]{l,black!10} BRH & Present & 17.70\% \\
 & Not Present & 83.30\% \\
\hline
\SetCell[c=1]{l,black!10} PNA & Present & 35.87\% \\
 & Not Present & 64.13\% \\
\hline
\SetCell[c=1]{l,black!10} RF & Present & 19.05\% \\
 & Not Present & 80.95\% \\
\hline
\end{tblr}
\label{tab: cat summ}
\end{table*}

\section{Methodology}
\label{sec:methodology}

\subsection{Preliminaries and related work}
Consider a survival response variable $T$ and a set of covariates $\boldsymbol{X} = (\boldsymbol{U},\boldsymbol{V})$, where $\boldsymbol{U}$ denotes a $p-$dimensional vector of continuous variables and $\boldsymbol{V}$ represents a $q-$dimensional vector of categorical variables. As is customary in survival analysis, we assume that the response $T$ may be right-censored. Therefore, 
the target variable is defined as $Y = \min\{T, C\}$, where $C$ is a nonnegative random variable that is independent of $T$ and represents the censoring mechanism. Furthermore, we observe a censoring indicator $\delta$, which is equal to $1$ if $T$ is observed and $0$ otherwise. 
We consider $\boldsymbol{X}$ and $Y$ defined in a finite space $\boldsymbol{\Omega}$ with values in $\mathbb{R}^{(p + q)} \times \mathbb{R}^+$, which is assumed to be partitioned into $G$ clusters denoted $\boldsymbol{\Omega}_1,\ldots, \boldsymbol{\Omega}_G$.
Given this setup, the joint probability across the clusters can be factorized as follows:
\begin{equation}
\label{eq:joint density}
p(Y, \boldsymbol{X}; \boldsymbol{\psi}) = \sum_{g= 1}^{G} \tau_{g}
p(y|\boldsymbol{x}; \boldsymbol{\gamma}_g, \boldsymbol{\beta}_g, \theta_g) \phi(\boldsymbol{u}; \boldsymbol{\mu}_g, \boldsymbol{\Sigma}_g) \xi(\boldsymbol{v}; \boldsymbol{\pi}_g), 
\end{equation}
where $\tau_g$ represents the positive mixing weights that sum to $1$, $p(y|\boldsymbol{x}; \boldsymbol{\gamma}_g, \boldsymbol{\beta}_g, \theta_g)$ denotes the conditional density of $Y | \boldsymbol{X}$ for the $g$-th component, whose full specification is provided in the next subsection, and the remaining terms correspond to the marginal densities of the covariates. In detail, we model continuous features $\mathbf{U}$ using a multivariate Gaussian distribution $\phi(\cdot;\boldsymbol{\mu}_g,\boldsymbol{\Sigma}_g)$, with cluster-wise different mean vectors $\boldsymbol{\mu}_g$ and covariance matrices $\boldsymbol{\Sigma}_g$. The density $\xi(\cdot;\bm{\pi}_{g})$ of the $q$ categorical covariates in $\mathbf{V}$, each potentially possessing a different number of categories, is given by the product of $q$ independent multinomial distributions  with cluster-wise different parameter for event probabilities $\boldsymbol{\pi}_{g}$, as in \cite{Ingrassia2015} and \cite{Berta2019}. Lastly, we use $\boldsymbol{\psi}=\{ \boldsymbol{\beta}_g, \theta_g, \boldsymbol{\gamma}_g, \tau_g, \boldsymbol{\mu}_g, \boldsymbol{\Sigma}_g,  \boldsymbol{\pi}_g \}_{g = 1}^G$ to denote the complete set of model parameters to be estimated.

The joint distribution presented in Equation \eqref{eq:joint density} defines a general Cluster-Weighted Model (CWM) framework, allowing the specification of various modeling approaches depending on the choice of the conditional density $p(y|\boldsymbol{x}; \boldsymbol{\gamma}_g, \boldsymbol{\beta}_g, \theta_g)$. Specifically, linear Gaussian CWMs and generalized linear CWMs arise when the conditional distributions are assumed to belong to the exponential family \citep{Ingrassia2012, Ingrassia2015}.
When data exhibit a hierarchical structure, multilevel linear Gaussian CWMs and multilevel generalized linear CWMs have been proposed in \cite{Berta2016} and \cite{Berta2019}, respectively. A recent extension of the latter, which incorporates dependencies among dichotomous covariates through the Ising model,  has been introduced by \cite{Caldera2025}.

Motivated by the clinical context described in Section \ref{subsec:prob_sett_and_data_descr}, we extend the family of multilevel Cluster-Weighted Models to accommodate time-to-event responses by incorporating a parametric frailty term into the specification of the conditional density. The detailed formulation of the model and the resulting likelihood function are presented in sections \ref{subsec:frailty_surv_model} and \ref{subsec:model_likelihood}. 

\subsection{On the specification of the likelihood term for the survival response}
\label{subsec:frailty_surv_model}
To clarify the modeling structure, we distinguish between clusters and groups, which play different roles in our framework. The clusters ($g = 1,\ldots,G$) are latent subpopulations identified by the mixture model, each characterized by specific distributional parameters and potentially distinct survival mechanisms. In contrast, the groups ($j = 1,\ldots,J$) refer to observed higher-level units, such as hospitals, clinical centers, or other natural aggregations in the data, within which individual observations are nested.
Building on this distinction, we now outline the proposed framework for extending the conditional density $p(y|\boldsymbol{x}; \boldsymbol{\gamma}_g, \boldsymbol{\beta}_g, \theta_g)$ in Equation \eqref{eq:joint density} to accommodate multilevel survival responses. Start by considering hierarchical time-to-event data, for which each statistical unit $i$, $i=1,\ldots, n_j$, within the group $j$, $j=1,\ldots,J$, is identified by the triplet $(y_{ij}, \delta_{ij}, \boldsymbol{x}_{ij})$, where:
\begin{itemize} 
    \item $y_{ij} = \min\{t_{ij}, c_{ij}\}$, being $t_{ij}$ the survival time and $c_{ij}$ the censoring time for individual $i$ in group $j$;
    \item $\delta_{ij} = \mathbbm{1}\{ t_{ij} < c_{ij} \}$ represents the event indicator for observation $i$ in group $j$;
    \item $\boldsymbol{x}_{ij} = (\boldsymbol{u}_{ij}, \boldsymbol{v}_{ij})$ denotes the vector of covariates with $\boldsymbol{u}_{ij}$ and $\boldsymbol{v}_{ij}$ indicating the subset of continuous and categorical predictors for the $ij-$th observation, respectively.
\end{itemize}
For each latent cluster $g$, we adopt a shared frailty model specified in terms of conditional hazard \citep{duchateau2008frailty, Munda2012}:

\begin{align}
\label{eq: dens haz}
    h(y_{ij} | m_{jg},\boldsymbol{x}_{ij}; \boldsymbol{\gamma}_g, \boldsymbol{\beta}_g) = h_0(y_{ij}; \boldsymbol{\gamma}_g) m_{jg} \exp\{\boldsymbol{x}_{ij}^T \, \boldsymbol{\beta}_g \},
\end{align}
where $\boldsymbol{\beta}_g$ is a vector of covariate coefficients and $m_{jg}$ is the shared frailty associated with group $j$ and cluster $g$, for $j=1,\ldots,J$ and $g=1,\ldots,G$, assumed to be independent random variables with density functions $f_M(\cdot;\theta_g)$. Here, $\theta_g$ is a cluster-specific parameter that quantifies the variability of the frailty within cluster $g$. The frailty term operates at the group level within each latent cluster. Specifically, for each cluster $g$, a shared frailty term $m_{jg}$ is assigned to each group $j$, capturing unobserved heterogeneity between groups within that cluster. As the frailty is both group and cluster specific, a given group $j$ is associated with distinct frailty terms $m_{jg}$ across the $G$ clusters, allowing for cluster-dependent group-level effects. Lastly, the term $h_0(\cdot; \boldsymbol{\gamma}_g)$ represents the baseline hazard function, parameterized by the vector of parameters $\boldsymbol{\gamma}_g$. Its specification depends on the chosen parametric distribution, with the Weibull being the most common choice for modeling the baseline. Alternative options considered in this study include the Exponential, Gompertz, and Lognormal distributions, as reported in Table~\ref{tab:bas_haz}. Further, define the (conditional) cumulative hazard function as follows:
\begin{equation}
\label{eq: cum haz}
    H(y_{ij}|m_{jg},\boldsymbol{x}_{ij}; \boldsymbol{\gamma}_g, \boldsymbol{\beta}_g) = \int_{0}^{y_{ij}} h(s | m_{jg},\boldsymbol{x}_{ij}; \boldsymbol{\gamma}_g, \boldsymbol{\beta}_g) ds= 
    m_{jg} \exp\{\boldsymbol{x}_{ij}^T \, \boldsymbol{\beta}_g \} H_0(y_{ij}; \boldsymbol{\gamma}_g).
\end{equation}

Starting from Equations \eqref{eq: dens haz} and \eqref{eq: cum haz} and assuming conditional independence, we write the conditional likelihood of the observations in group $j$ assigned to cluster $g$ as follows:
\begin{equation}
\prod_{i \in R_{jg}}h(y_{ij} | m_{jg},\boldsymbol{x}_{ij}; \boldsymbol{\gamma}_g, \boldsymbol{\beta}_g)^{\delta_{ij}} S(y_{ij} | m_{jg},\boldsymbol{x}_{ij}; \boldsymbol{\gamma}_g, \boldsymbol{\beta}_g),
\end{equation}
where $S(y_{ij} | m_{jg}, \boldsymbol{x}_{ij}; \boldsymbol{\gamma}_g, \boldsymbol{\beta}_g)= \exp\{-H(y_{ij}|m_{jg}, \boldsymbol{x}_{ij}; \boldsymbol{\gamma}_g, \boldsymbol{\beta}_g)\}$, see e.g., \cite{klein2006survival}, and $R_{jg}$ contains the indexes of the observations in group $j$ assigned to cluster $g$. 
\begin{table*}[t]
\caption{Parametric distributions considered in the specification of the conditional densities for the survival response. Here, $\phi$ and $\Phi$ respectively indicate the probability density and the cumulative distribution of a standard Gaussian, while  $\boldsymbol{\gamma}$ is used as a generic notation to represent the set of parameters for the different baselines. Table adapted from \cite{Munda2012}. }
\renewcommand{\arraystretch}{1.8}
\centering
\large
\begin{tblr}{l c c c}
\hline
  \textbf{Distribution} & $h_0(t; \boldsymbol{\gamma})$ & $H_0(t;\boldsymbol{\gamma}) = \int_0^t h_0(s;\boldsymbol{\gamma}) ds$ & \textbf{Parameter Space} $\boldsymbol{\gamma}$\\ 
 \hline \hline
Exponential & $\lambda$ & $\lambda t$ & $\lambda > 0$ \\
\hline
Weibull & $\lambda \rho t^{\rho - 1}$ & $\lambda t^{\rho}$ & $\lambda, \rho > 0$ \\
\hline
Gompertz & $\lambda \exp(\rho t)$ & $\frac{\lambda}{\rho} (\exp(\rho t) - 1)$ & $\lambda, \rho > 0$ \\
\hline
Lognormal & $\frac{\phi\bigl(\frac{log(t) - \eta}{\nu}\bigr)}{\nu t \bigl[ 1 - \Phi\bigl(\frac{log(t) - \eta}{\nu}\bigr) \bigr]}$ & $-\log\bigl[ 1 - \Phi\bigl(\frac{log(t) - \eta}{\nu}\bigr) \bigr]$ & $\eta \in \mathbbm{R}, \nu > 0$  \\
\hline
\end{tblr}
\label{tab:bas_haz}
\end{table*}
As a final step, to obtain the likelihood contribution for the survival response used in the CWM specification of Equation \eqref{eq:joint density}, the unobservable random effects $m_{jg}$ must be integrated out with respect to the marginal density of the frailty, resulting in the following expression:

\begin{equation}
L_{jg}^S(\boldsymbol{\gamma}_g, \boldsymbol{\beta}_g,\theta_g)=\int_{0}^{+\infty} \prod_{i \in R_{jg}}h(y_{ij} | m_{jg},\boldsymbol{x}_{ij}; \boldsymbol{\gamma}_g, \boldsymbol{\beta}_g)^{\delta_{ij}} S(y_{ij} | m_{jg},\boldsymbol{x}_{ij}; \boldsymbol{\gamma}_g, \boldsymbol{\beta}_g) f_M(m_{jg};\theta_g) dm_{jg}.
\end{equation}
Given a sample of $N = \sum_{j=1}^{J} n_j$ observation triplets $(\mathbf{x}_{ij}, t_{ij}, \delta_{ij})$, a Classification Maximum Likelihood criterion is formulated to estimate the model parameters, as detailed in the next subsection.

\subsection{Objective Function of the Model}
\label{subsec:model_likelihood}
In defining the objective function of the model, we adopt a Classification Maximum Likelihood (CML) approach, treating the latent assignment of observations to the mixture components as unknown parameters \citep{Bryant1978, Celeux1992}. There are two key reasons for this choice. First, the primary objective of the motivating application is to identify clusters (also referred to as profiles) of patients. The CML framework is naturally aligned with this goal, as standard clustering algorithms can be viewed as specific instances of CML criteria \citep[see, e.g., ][and references therein]{Jain1988, Celeux1992, Garcia-Escudero2008}. Second, the ``all-or-nothing'' assignment inherent in the CML framework simplifies computation, enabling the use of readily available routines to independently maximize the contributions of the $G$ components (see Section \ref{sec:m-step}). In detail, based on the general CWM specification outlined in Equation \eqref{eq:joint density}, and incorporating the contribution of the survival response derived in the previous section, the resulting likelihood takes the following form:
\begin{equation}
\label{eq:lik}
  L(\boldsymbol{\psi}) = \prod_{g=1}^G\prod_{j=1}^JL_{jg}^S(\boldsymbol{\gamma}_g, \boldsymbol{\beta}_g,\theta_g)
  \prod_{i \in R_{jg}}\tau_g\phi(\boldsymbol{u}_{ij}; \boldsymbol{\mu}_g, \boldsymbol{\Sigma}_g) \xi(\boldsymbol{v}_{ij}; \boldsymbol{\pi}_g),
    \end{equation}
    
where the conditional density of $Y | \boldsymbol{X}$ for the $g$-th component now explicitly corresponds to the likelihood of a parametric frailty model, with the frailties integrated out by averaging the conditional likelihood over the frailty distribution \citep{Munda2012}. Ultimately, the overall objective function of the model can be expressed in terms of the following classification log-likelihood:

\begin{align}
\begin{split}
\label{eq: compl loglik 5}
    \ell\left(\boldsymbol{\psi}\right)= \sum_{g=1}^G&\left\{\sum_{j=1}^J \left[ \sum_{i \in R_{jg}} \delta_{ij}\left(\log{h_0(y_{ij}; \boldsymbol{\gamma}_g)}+ \mathbf{x}_{ij}^{'}\boldsymbol{\beta}_g \right)\right. \right.+\\
    &\left. \log{\left[ (-1)^{d_{jg}}\mathcal{L}^{(d_{jg})}\left( \sum_{i \in R_{jg}} H_0(y_{ij}; \boldsymbol{\gamma}_g)\exp{\left(\mathbf{x}_{ij}^{'}\boldsymbol{\beta}_g \right)};\theta_g\right)\right]}\right] +\\
   &\left. \sum_{j=1}^J\sum_{i \in R_{jg}} \left(  \log{\tau_g} + \log{\phi(\mathbf{u}_{ij}; \boldsymbol{\mu}_g,\boldsymbol{\Sigma}_g)}+ \log{\xi(\boldsymbol{v}_{ij}; \boldsymbol{\pi}_g)} \right) \right\},
\end{split}    
\end{align}
where, $d_{jg} = \sum_{i \in R_{jg}} \delta_{ij}$ represents the number of events for observations belonging to group $j$ and assigned to cluster $g$, and $\mathcal{L}^{(q)}(\cdot)$ denotes the $q$-th derivative of the Laplace transform of the frailty distribution. The complete derivation of the objective function in Equation \eqref{eq: compl loglik 5} is provided in Supplement A. Direct maximization of Equation \eqref{eq: compl loglik 5} poses a complex optimization problem. To address this, we propose two EM-based algorithms for parameter estimation: one incorporating a classification step (CEM algorithm), and the other relying on a stochastic step (SEM algorithm), as detailed in the next section.

 \subsection{Model Estimation}
\label{subsec:model_estimation}
Given the objective function of our method, expressed as the classification log-likelihood of Equation \eqref{eq: compl loglik 5}, we maximize it using variants of the classical Expectation-Maximization algorithm \citep{Dempster1977}. Specifically, we introduce a ``hard assignment'' phase between the E-step and the M-step of the standard EM procedure, implemented either through a classification step (C-step) based on the maximum a posteriori (MAP) principle, or a stochastic step (S-step) which simulates the partition according to the posterior probabilities obtained in the E-step. Depending on how the hard assignment is performed, this leads to the Classification EM (CEM) algorithm \citep{Celeux1992} or the Stochastic EM (SEM) algorithm \citep{celeux1985sem} for parameter estimation. The CEM algorithm shares the theoretical guarantees of the standard EM procedure, ensuring monotonicity of the objective function at each iteration and convergence to a stationary point. However, it is highly sensitive to initialization and may become trapped in local optima. Such drawbacks are not shared by the SEM algorithm, which was specifically developed to overcome these limitations, albeit at the expense of losing the ascent property and having more complex convergence behavior \citep{Nielsen2000}. The SEM algorithm has also been successfully applied in the literature to fit mixture models with censored data \citep{chauveau1995stochastic, Bordes2016}.

Despite their differing advantages and limitations, both the CEM and SEM procedures share a common characteristic: they generate a hard partition at each algorithm iteration. This significantly simplifies the M-step by enabling independent maximization of each component's contribution, considering only the units assigned to that component in the current iteration. In practice, such a characteristic enables the use of standard maximum likelihood estimates for homogeneous populations, thereby leveraging existing computational routines. This proves particularly advantageous when estimating the parametric frailty term in the conditional density of our CWM specification. Detailed algorithmic steps are described in the following subsections.

\subsubsection{E-step}
As is standard in mixture models \citep[see, for instance,][]{bouveyron2019model}, the $(k + 1)$-th iteration of the Expectation step involves computing the posterior probability that observation $i$ from group $j$ belongs to cluster $g$, given the parameter estimates obtained in the previous step. To formalize this, we introduce assignment indicators $z_{ijg}$, where $z_{ijg} = 1$ if observation $i$ from hospital $j$ is assigned to the $g$-th component, and $z_{ijg} = 0$ otherwise, for $i = 1, \ldots, n_j$, $j = 1, \ldots, J$, and $g = 1, \ldots, G$. Then, the estimated a posteriori probabilities are routinely updated as follows:

\begin{equation} \label{eq:estep}  
    \hat{z}_{ijg}^{(k+1)}= \frac{\hat{\tau}^{(k)}_g p(y_{ij}|\boldsymbol{x}_{ij}; \hat{\boldsymbol{\gamma}}^{(k)}_g, \hat{\boldsymbol{\beta}}^{(k)}_g, \hat{\theta}^{(k)}_g)\phi(\boldsymbol{u}_{ij}; \hat{\boldsymbol{\mu}}^{(k)}_g, \hat{\boldsymbol{\Sigma}}^{(k)}_g) \xi(\boldsymbol{v}_{ij}; \hat{\boldsymbol{\pi}}^{(k)}_g)}{\sum_{c=1}^G\hat{\tau}^{(k)}_c p(y_{ij}|\boldsymbol{x}_{ij}; \hat{\boldsymbol{\gamma}}^{(k)}_c, \hat{\boldsymbol{\beta}}^{(k)}_c, \hat{\theta}^{(k)}_c)\phi(\boldsymbol{u}_{ij}; \hat{\boldsymbol{\mu}}^{(k)}_c, \hat{\boldsymbol{\Sigma}}^{(k)}_c) \xi(\boldsymbol{v}_{ij}; \hat{\boldsymbol{\pi}}^{(k)}_c)},
\end{equation}
where the superscript $k$ denotes the parameter estimates obtained in the previous EM iteration.
\subsubsection{Hard assignment via C-step or S-step}
Starting from the soft assignments computed in Equation \eqref{eq:estep}, we employ either the MAP rule or a stochastic sampling procedure to hard-assign the units to the $G$ clusters. Specifically, the hard assignment update in the CEM algorithm  requires the following Classification step:

\begin{equation} \label{eq:cstep}
\tilde{z}^{(k+1)}_{ijg}=\begin{cases} 1 & \text{if } g=\operatorname*{argmax}_{c \in \{1,\ldots,G\}} \ \hat{z}_{ijc}^{(k+1)} \\ 0  & \text{otherwise} \end{cases}.
\end{equation}
When the SEM algorithm is used instead to maximize Equation \eqref{eq: compl loglik 5}, the hard assignment update involves the following Stochastic step:
\begin{equation} \label{eq:sstep}
\tilde{z}^{(k+1)}_{ijg} \sim Multinomial(1, \hat{\boldsymbol{z}}_{ij}^{(k+1)}),
\end{equation}
where the allocation is obtained by sampling from a multinomial distribution with event probabilities $$\hat{\boldsymbol{z}}_{ij}^{(k+1)}=\left(\hat{z}_{ij1}^{(k+1)},\ldots \hat{z}_{ijG}^{(k+1)}\right).$$
Whether using the C-step in Equation~\eqref{eq:cstep} or the S-step in Equation~\eqref{eq:sstep}, the units are partitioned into $G$ clusters whose sample size is then computed as:

$$
\hat{n}^{(k+1)}_g=\sum_{j=1}^J\sum_{i=1}^{n_j} \tilde{z}^{(k+1)}_{ijg}, \quad g=1,\ldots G.
$$
Note that there exists a one-to-one correspondence between the index notation $R_{jg}$, as introduced in Sections \ref{subsec:frailty_surv_model} and \ref{subsec:model_likelihood}, and the assignment indicators $z_{ijg}$. Specifically, at each iteration of the algorithm, the sets $R_{jg}$ are updated to include the indices of all observations in hospital j for which $\tilde{z}^{(k+1)}_{ijg} = 1$.

\subsubsection{M-step} \label{sec:m-step}
Thanks to the partition induced by the hard assignment described in the previous section, the M-step reduces to separately maximizing the likelihood contributions of $G$ distinct subpopulations, each with a sample size equal to $\hat{n}^{(k+1)}_g$, for $g = 1, \ldots, G$. Specifically, traditional maximum likelihood estimates can be easily obtained for parameters $\{\tau_g, \boldsymbol{\mu}_g, \boldsymbol{\Sigma}_g, \boldsymbol{\pi}_g \}_{g = 1}^G$, which govern mixing proportions and marginal distributions of both continuous and categorical covariates. Detailed derivations of these formulas can be found, for instance, in the supplementary material of \cite{Caldera2025}. Maximizing the likelihood term related to the survival response is more complex and depends heavily on the specified parametric forms. To address this challenge, we utilize the \texttt{parfm R} package \citep{Munda2012}, which provides a unified framework for fitting parametric frailty models. This approach offers two key advantages: it ensures computational efficiency and retains flexibility by allowing users to select from a wide range of parametric distributions for both the baseline hazard and frailty term. For a comprehensive discussion of the underlying maximization strategies, we refer the reader to \cite{Munda2012}.

\subsection{Additional Details on the Estimation Procedure}
\label{subsubsec:model_additional_details} 

We hereafter discuss several practical considerations related to the estimation procedure.

\begin{itemize}
    \item \textit{Initialization}: the initialization of deterministic algorithms is always a critical step, and this is especially true when dealing with time-to-event data. To date, only practical data-driven heuristics have been proposed to initialize EM algorithms in mixture models for right-censored lifetime data \citep{Bordes2016}. However, in our approach, we can leverage the postulated differences in the covariate distributions to construct an initial partition. Specifically, we employ k-prototypes, an extension of k-means designed for data sets containing both continuous and categorical variables \citep{huang1998extensions}. Although other established alternatives such as multiple random initializations are possible, numerical experiments on both real and synthetic data suggest that the aforementioned strategy effectively guides the algorithms toward a stable convergence path.
    \item \textit{Convergence}: for the CEM algorithm, convergence is assumed when the relative difference in the objective function defined in Equation \eqref{eq: compl loglik 5} between two consecutive iterations falls below a threshold $\varepsilon$. In our analyses, we set $\varepsilon = 10^{-5}$. More sophisticated convergence diagnostics are required to assess the convergence of the SEM algorithm, as it does not guarantee a monotonically increasing objective function. Specifically, the final values are defined as the ergodic mean of the sequence of parameter estimates across iterations, as proposed by \citet{Bordes2016} based on the asymptotic properties established in \citet{Nielsen2000}.
    \item \textit{Model selection}: the Bayesian Information Criterion \citep[BIC;][]{Schwarz1978} is used to select, in a data-driven manner, the number of clusters $G$ as well as the parametric form of both the baseline and frailty distributions. The criterion reads: 
    \begin{align}
   \label{eq:BIC}
  \text{BIC} &= 2 \cdot \ell(\hat{\boldsymbol{\psi}}) - d \cdot \ln(N),
\end{align}

where $\ell(\hat{\boldsymbol{\psi}})$ denotes the maximized log-likelihood, \( N \) is the sample size and \( d \) represents the total number of parameters: 
$$
d = G( 1 + m ) +G\left( \frac{p(p+3)}{2} \right)+G\sum_{r = 1}^{q} (k_r - 1) +Gb+Gf+G - 1.
$$
In details, \( k_r \) defines the number of categories associated with the \( r \)-th categorical variable, \( m \) is the number of covariates included in the regression component of the shared frailty survival model, \( b \) is the number of parameters of the chosen baseline distribution, and \( f \) is the number of parameters of the frailty distribution. According to the definition of Equation \eqref{eq:BIC}, the best model is the one with the highest BIC.
    \item \textit{Implementation}: routines for implementing the proposed methodology through both CEM and SEM algorithms have been developed in \texttt{R} \citep{RCoreTeam}, with the source code freely available at https://github.com/AndreaCappozzo/mixparfmCWM. As discussed in Section \ref{sec:m-step}, maximization of the survival term relies on the \texttt{parfm R} package \citep{Munda2012}. However, to accommodate the specific challenges posed by our framework, the original \texttt{parfm} routines have been slightly modified and extended. An enriched version of  \texttt{parfm}, available at https://github.com/AndreaCappozzo/parfm, is required for the devised \texttt{mixparfmCWM R} package to function correctly.
\end{itemize}

\section{Application and Results}
\label{sec:applic_and_results}
We apply the proposed methodology to the Lombardy region dataset, described in Section~\ref{subsec:prob_sett_and_data_descr}, with the aim of profiling patients and modeling their hazard of death based on cluster-specific respiratory conditions and hospital effect. Both the CEM and SEM algorithms were considered for model fitting, yielding virtually identical parameter estimates. Consequently, the results presented in the following analysis are based on the CEM algorithm.

\subsection{Model Setting}
\label{subsubsec:model_setting_app}
The variables Age, Gender, MCS, COPD, and BRH represent pre-existing conditions prior to the onset of COVID-19 pathology. Consequently, they are incorporated into the marginal distribution and modeled as random covariates within the adopted CWM framework. Specifically, continuous variables (MCS and Age) are modeled jointly using a bivariate Gaussian distribution with cluster-specific parameters $\{\boldsymbol{\mu}_g, \boldsymbol{\Sigma}_g\}$. The variables Gender, COPD, and BRH are treated as independent binary categorical variables, with one parameter vector $\boldsymbol{\pi}_g$ estimated for each cluster. These pre-existing condition variables are excluded from the set of covariates in the parametric frailty model term. In contrast, the variables PNA and RF, which generally manifest as complications arising from COVID-19 infection \citep{du2020predictors,zhou2020clinical}, are included as explanatory variables in the survival regression term. This modeling strategy is designed to allow the pre-existing conditions (Age, Gender, MCS, COPD, and BRH) to primarily inform the identification of latent clusters (i.e., patient profiles), while the influence of respiratory complications is adjusted through their effect on cluster-specific survival outcomes. Accordingly,  each cluster $g$ is associated with a vector of regression coefficients, $\boldsymbol{\beta}_g$, capturing the impact of PNA and RF on survival.

We fit the model introduced in Section \ref{sec:methodology} by varying \( G \) over the set \( \{1, 2, 3, 4, 5\} \) and considering four baseline hazard distributions: Exponential, Weibull, Lognormal, and Gompertz. For the frailty term, we assume a Gamma distribution with a fixed mean of 1 and an unknown variance, denoted by \(\theta_g\). Following the procedure outlined in Section~\ref{subsubsec:model_additional_details}, the model is fitted 20 times for each combination of \( G \) and baseline hazard distribution. To initialize the algorithm, we use k-prototypes clustering based on the five variables modeled as random: MCS, Age, Gender, COPD, and BRH. The best BIC value across the 20 runs for each combination of the number of clusters \( G \) and baseline hazard distribution is reported in Table~\ref{tab:bic_comp}. The optimal model, identified by the highest BIC, corresponds to \( G = 3 \) with a Lognormal baseline distribution.

\begin{table*}[htbp]
\caption{Comparison of BIC values across all combinations of number of clusters $G$ and Baseline Hazard Distributions. The Model with the highest BIC is highlighted in bold. 
}
\renewcommand{\arraystretch}{1.6}
\centering
\begin{tblr}{c c c c c c}
\hline
  \textbf{Baseline Distribution} & $\boldsymbol{G = 1}$ & $\boldsymbol{G = 2}$ &  $\boldsymbol{G = 3}$  &  $\boldsymbol{G = 4}$  &  $\boldsymbol{G = 5}$   \\ 
 \hline \hline
Exponential & $-54198.20$ & $-54155.42$ & $-53297.05$ & $-53386.16$ & $-53313.37$ \\
\hline
Weibull & $-53763.28$ & $-53745.27$ & $-52891.27$ & $-52982.42$ & $-52947.69$ \\
\hline
Lognormal & $-53372.23$ & $-53355.70$ & $\boldsymbol{-52594.26}$ & $-52610.39$ & $-52635.10$ \\
\hline
Gompertz & $-54206.23$ & $-54169.07$ & $-53322.20$ & $-53427.78$ & $-53408.48$ \\
\hline
\end{tblr}
\label{tab:bic_comp}
\end{table*}

\subsection{Results}
\label{subsubsec:results_app}
\begin{figure}[htpb]
\centerline{%
\includegraphics[width=\textwidth]{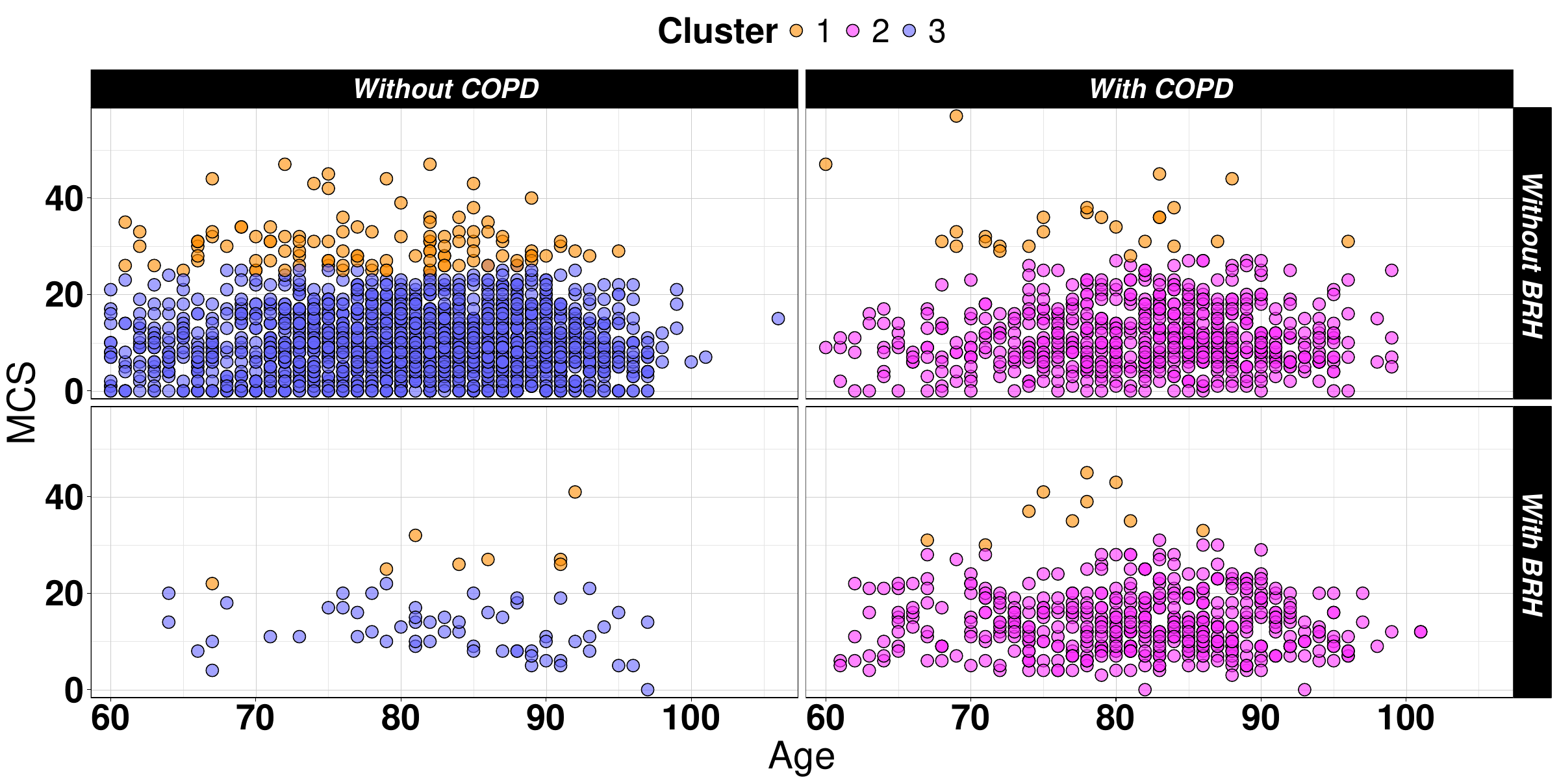}}
\caption{Scatterplots of patients in the Age–MCS feature space colored according to the three estimated clusters; stratified by Presence or Absence of COPD and BRH.}
\label{fig:clusters_plot}
\end{figure}

In this section, we present the results of the best model selected using BIC, focusing on clusters description, interpretation of survival parameters, and analysis of frailty effects. Figure~\ref{fig:clusters_plot} illustrates the partition of patients into the 3 clusters within the Age-MCS space, distinguishing between patients with and without COPD and BRH. The upper portion of Table~\ref{tab:app_parameters} provides a summary of the parameters associated with continuous and categorical variables, as well as the number of patients in each cluster. The lower portion of the table presents the estimated values and significance of the cluster-specific survival parameters, including fixed effects, the parameters that specify the Lognormal baseline hazard distribution and the estimated variance of the $\text{Gamma}$ distribution used to model the frailty term. 

\begin{figure}[htpb]
\centerline{%
\includegraphics[width=\textwidth]{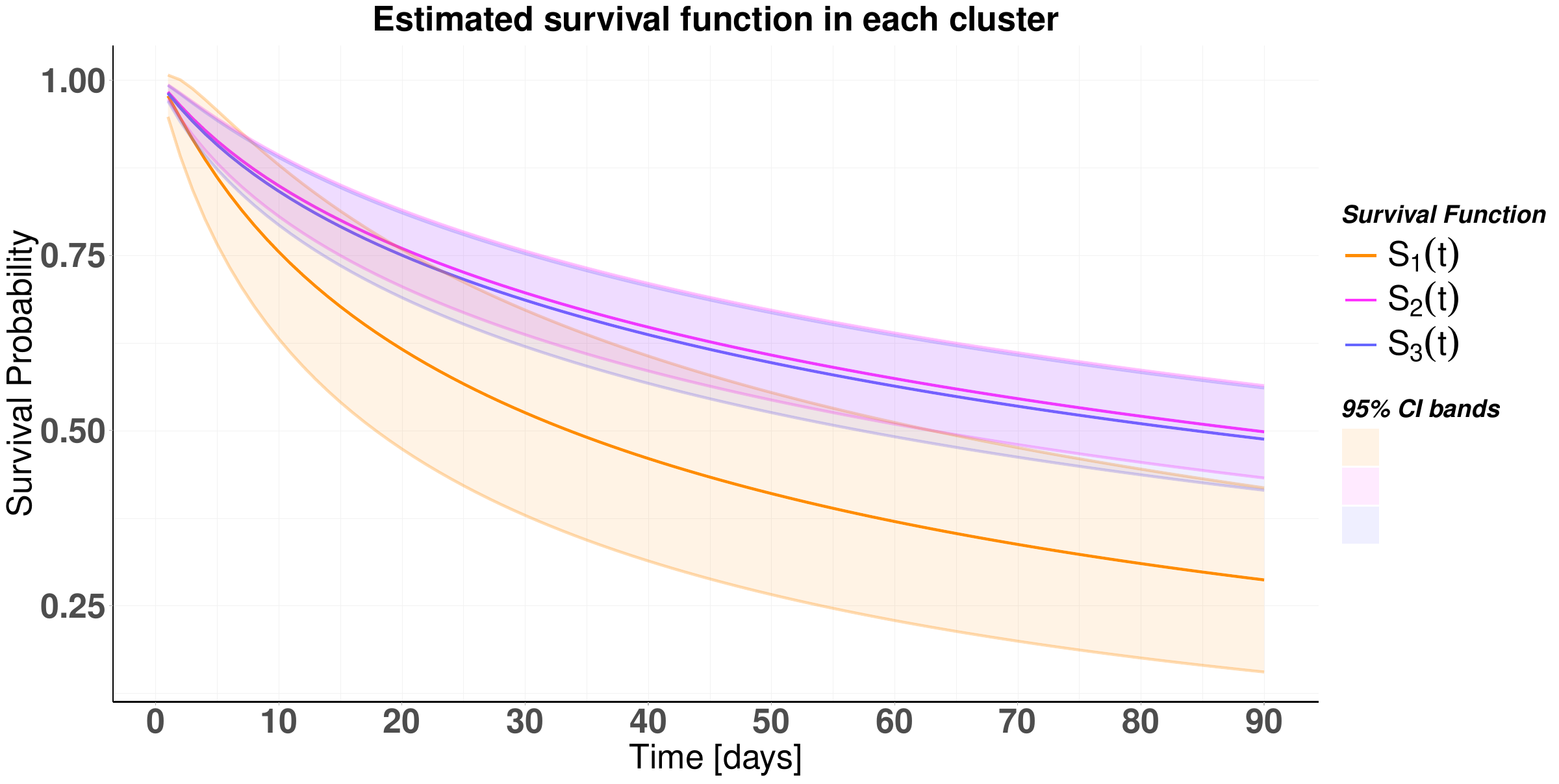}}
\caption{Estimated survival curves for the three clusters. The orange curve represents the baseline survival curve for Cluster 1, the violet curve corresponds to Cluster 2, and the blue curve represents Cluster 3. Curves refer to patients not affected by RF and PNA and with the frailty term fixed at its expected value, namely 1.Shaded bands indicate the 95\% confidence intervals for the survival curves.}
\label{fig:surv_per_clust}
\end{figure}

\begin{figure}[htpb]
    \centering
    \begin{subfigure}[b]{1\textwidth}
     \centering
        \includegraphics[width=0.8\textwidth]{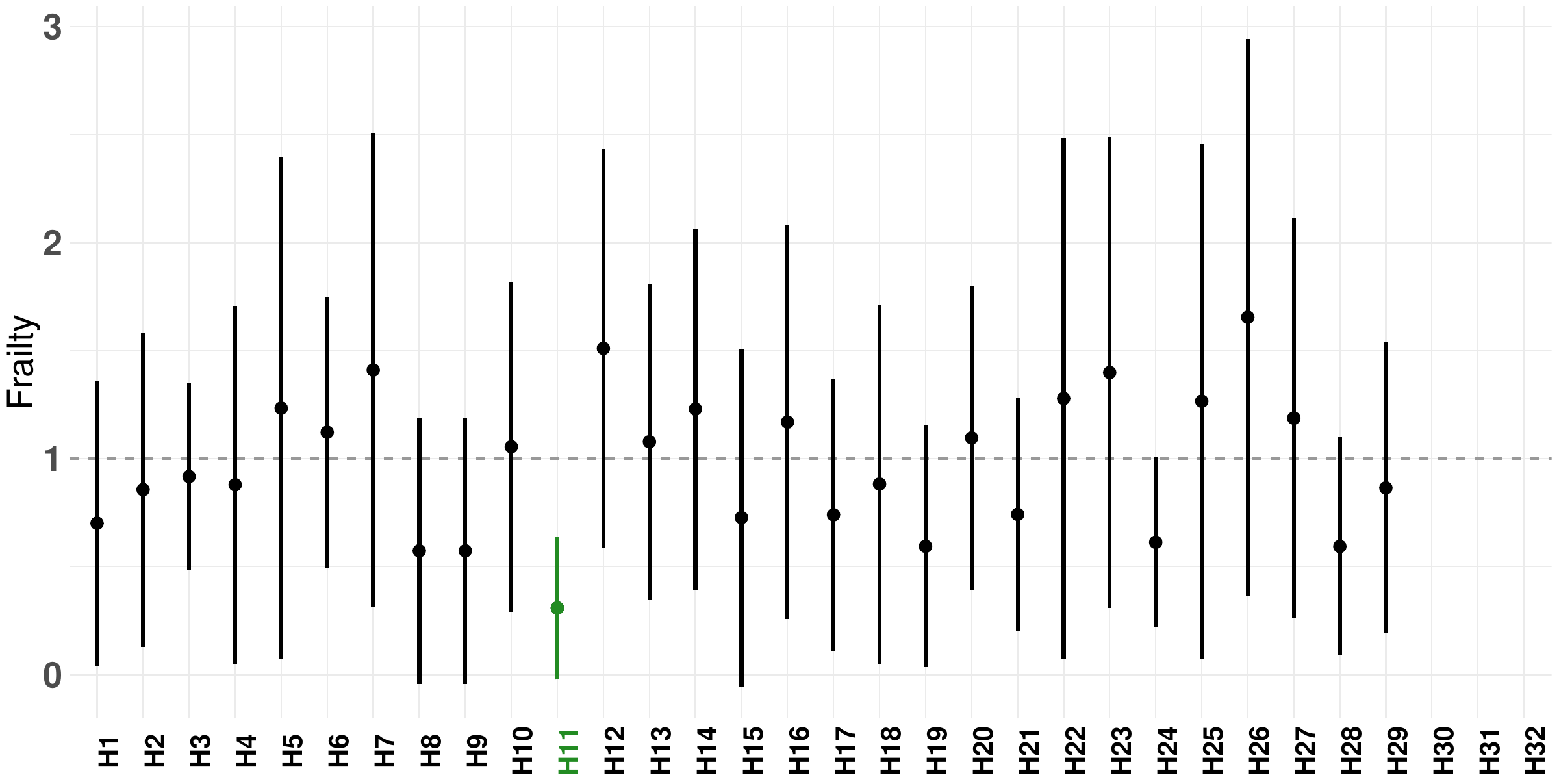}
        \caption{Estimated frailties for the 29 hospitals in Cluster 1. }
        \label{fig:Hosp1}
    \end{subfigure}
    \hfill
    \begin{subfigure}[b]{1\textwidth}
     \centering
        \includegraphics[width=0.8\textwidth]{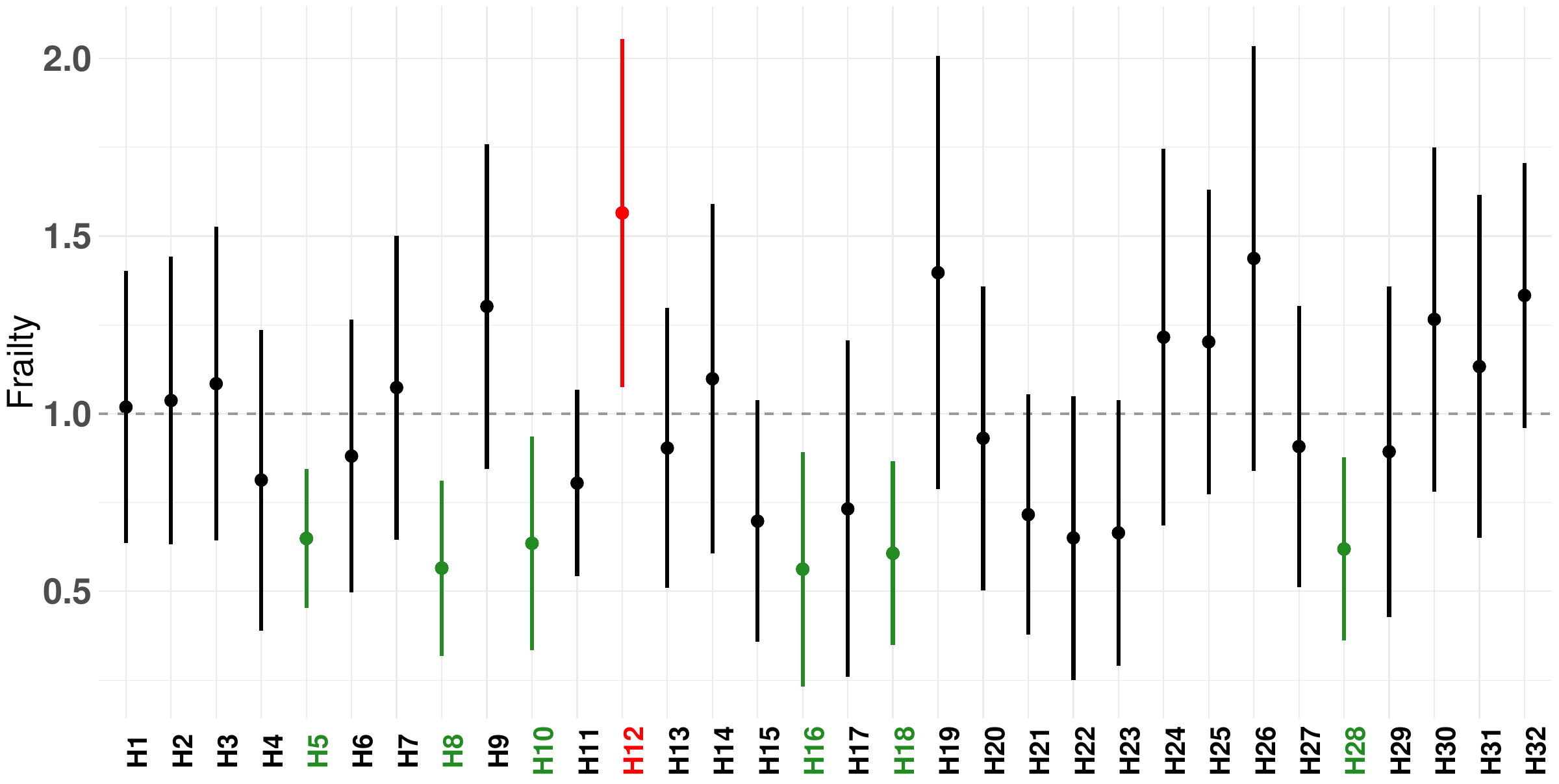} 
        \caption{Estimated frailties for the 32 hospitals in Cluster 2. }
        \label{fig:Hosp2}
    \end{subfigure}
    \hfill
    \begin{subfigure}[b]{1\textwidth}
     \centering
        \includegraphics[width=0.8\textwidth]{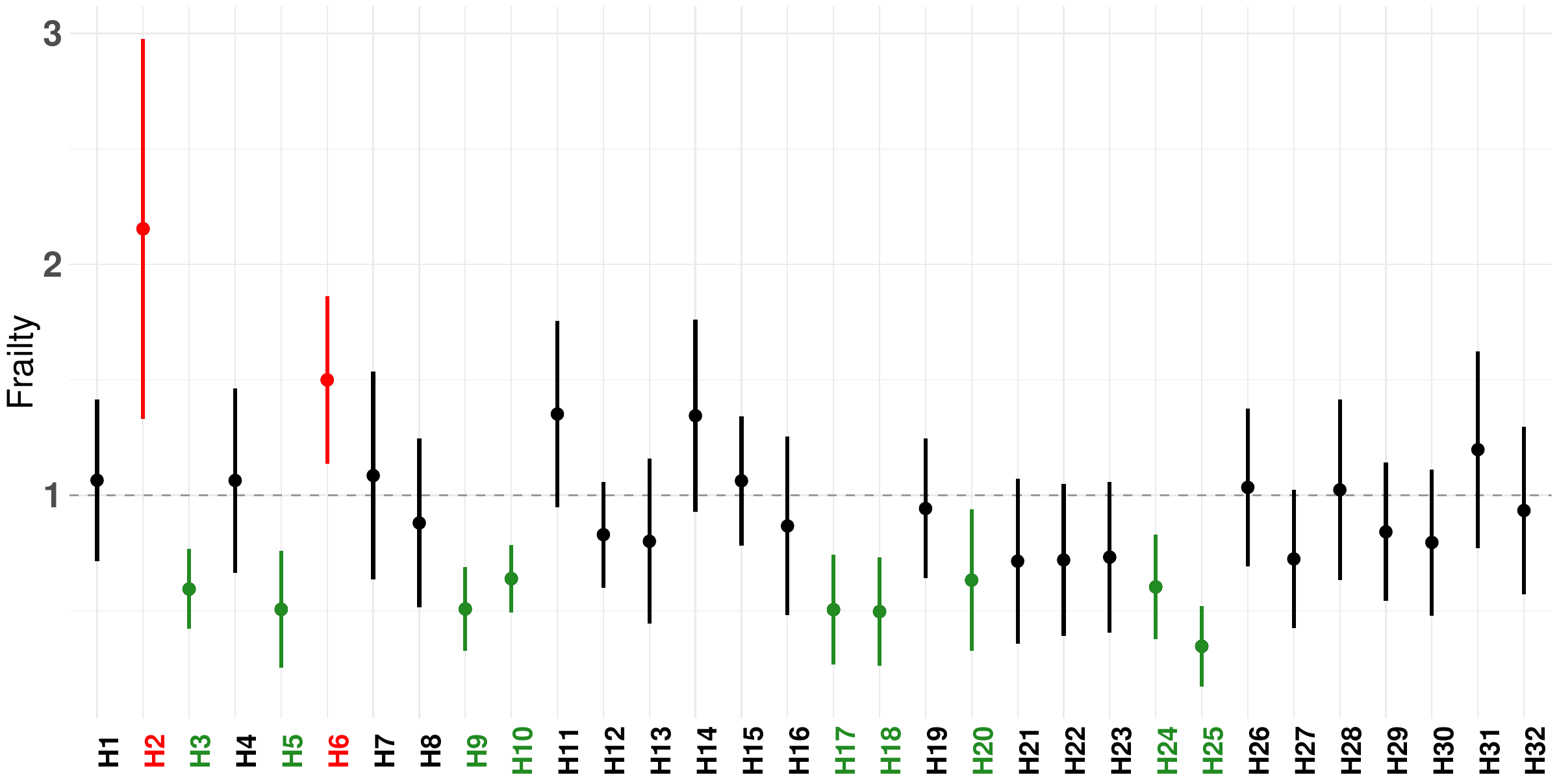}
        \caption{Estimated frailties for the 32 hospitals in Cluster 3.}
        \label{fig:Hosp3}
    \end{subfigure}
    \caption{Estimated frailties and their 95\% confidence intervals in the three clusters. Hospitals that significantly increase the hazard rate are highlighted in red, while those that decrease it are highlighted in green.}
    \label{fig:frailties_plot}
\end{figure}

In Figure~\ref{fig:surv_per_clust}, we represent the baseline survival function specific to each cluster, that corresponds to the estimated survival function for a patient not affected by RF and PNA and with the frailty term fixed at its expected value, namely 1. 
Given that a Lognormal distribution is considered for modeling the baseline hazard, the survival function \( S(t; \hat{\eta}_g, \hat{\nu}_g) \), for each cluster $g \in \{1,2,3\}$, is given by:
\[
S(t; \hat{\eta}_g, \hat{\nu}_g) = 1 - \Phi\left( \frac{\log(t) - \hat{\eta}_g}{\hat{\nu}_g} \right),
\]
where \( \hat{\eta}_g \) and \( \hat{\nu}_g \) are, for each cluster, the parameter estimates reported in the lower portion of Table~\ref{tab:app_parameters}. Additionally, Figure~\ref{fig:surv_per_clust} presents the corresponding 95\% confidence interval bands, computed using the delta method. Further details on the computation of these intervals are provided in Supplement C. In Figure~\ref{fig:app_hazard}, we represent the hazard rate over time for each cluster together with the effect of the statistically significant covariates. The estimated fixed effects \( \hat{\beta}_{\text{PNA}_g} \) and \( \hat{\beta}_{\text{RF}_g} \) represent the log-relative hazard for an individual in cluster $g$ with the specific disease relative to an individual in cluster $g$ without the disease. The interpretation of the coefficient is given in terms of the hazard ratio, which reflects the relative risk of the event occurring between the two levels of a binary covariate. A positive coefficient indicates that the hazard is higher for individuals with the condition compared to those without it, meaning the event is more likely to occur in the former group. In contrast, a negative coefficient suggests a lower hazard for individuals with the condition, indicating a protective effect. 
Figure~\ref{fig:frailties_plot} presents cluster-specific dot plots displaying the estimated hospital frailties along with their 95\% confidence intervals. These estimates capture the unobserved heterogeneity in survival outcomes across hospitals within each cluster, reflecting the extent to which individual facilities influence the baseline hazard due to unmeasured factors, such as differences in hospital practices, available resources, or other unobserved characteristics affecting patient survival. 
A joint examination of Figures~\ref{fig:surv_per_clust}, \ref{fig:app_hazard}, and \ref{fig:frailties_plot} enables the interpretation and characterization of the resulting clusters.

The first cluster (depicted in orange in the plots) includes 151 patients (4.9\% of the total cohort). The average age of this cluster of patients is 78.09 years, with a mean MCS of 31.79. The proportion of males exceeds that of females, as in the overall dataset. Indeed, all clusters exhibit a slightly higher proportion of male patients. With respect to respiratory conditions, 24\% of patients in this cluster are affected by COPD, whereas only 12\% present with BRH. In general, this cluster mainly includes individuals with a high burden of comorbidity but relatively low rates of BRH and COPD. This cluster exhibits the most unfavorable survival (see the orange curve in Figure~\ref{fig:surv_per_clust}), reflecting the lowest life expectancy and the highest hazard rate over the entire observation period (Figure~\ref{fig:app_hazard}). The effects of PNA and RF are not statistically significant within this cluster, suggesting that these conditions do not meaningfully influence the hazard rate for patients in this profile. This may be attributed to the already high burden of comorbidities characterizing the cluster, which likely attenuates the marginal impact of additional complications. With respect to the frailty term, there is no strong evidence of substantial heterogeneity across hospitals, as indicated by the estimated $\hat{\theta}$ reported in Table~\ref{tab:app_parameters} and the hospital-specific frailty estimates shown in Figure~\ref{fig:Hosp1}. Only one facility, Hospital H11, exhibits a notable protective effect for patients in this cluster. This further supports the interpretation that the severe baseline health conditions diminish the relative influence of the treating hospital, thereby reducing observable differences in survival outcomes across healthcare facilities. 

The second cluster (depicted in violet) includes 1071 patients (34.7\% of the total cohort). The average age of patients in this cluster is 81.79 years, with a mean MCS of 12.13. All patients in this cluster have COPD and 44\% of them also have BRH. This suggests a patient profile characterized by a very high prevalence of respiratory diseases and relatively fewer other comorbidities compared to the first cluster. The hazard rate for this cluster is lower over the entire time period (Figure~\ref{fig:app_hazard}), and the survival curve shows a higher life expectancy compared to the first cluster (Figure~\ref{fig:surv_per_clust}). The effect of PNA is significant, with an associated estimated coefficient of 0.235, suggesting that a patient in this cluster affected by PNA has a 23.5\% higher risk of death compared to a patient without PNA. The frailty effect is significant in this cluster, as shown in Table~\ref{tab:app_parameters}, suggesting heterogeneous effects of hospitals on patients survival. Specifically, hospitals H5, H8, H10, H16, H18, and H28 demonstrate a protective effect. In contrast, the confidence interval associated with hospital H12 lies entirely above 1, indicating a significantly increased hazard of death for patients in the second cluster treated at this facility (see Figure~\ref{fig:Hosp2}).

The third cluster (depicted in blue) comprises the largest proportion of the sample, including 1864 patients, which represents 60.4\% of the total cohort. The average age in this cluster is 81.38 years, with a mean MCS of 9.44. None of the patients in this cluster have COPD, and only 3\% have BRH. This cluster exhibits the lowest mean MCS and the lowest percentage of patients affected by COPD and BRH among the three clusters. Consequently, the patient profile in this cluster corresponds to the healthiest subgroup within the cohort. The survival curve and hazard rate over time closely resemble those of the second cluster (see Figures~\ref{fig:surv_per_clust} and~\ref{fig:app_hazard}). This similarity, despite the comparatively better health status of patients in the third cluster, particularly in terms of comorbidities and respiratory conditions, may be attributed to the prioritization of clinical monitoring and treatment for individuals with respiratory diseases, especially during the COVID-19 pandemic \citep{tiotiu2021impact, benfante2021prioritizing, giustivi2021respiratory}.
 Such prioritization likely contributed to comparable survival outcomes across these two otherwise distinct patient profiles. Notably, the effect of PNA is significant in this cluster, with an associated estimated coefficient of 0.167. This indicates that a patient in this cluster affected by PNA has a 16.7\% higher risk of death compared to a patient without PNA. Additionally, the frailty effect is also significant, as detailed in Table~\ref{tab:app_parameters}. Specifically, hospitals H3, H5, H9, H10, H17, H18, H20, H24, and H25 are associated to a decreased death hazard for patients treated there, whereas hospitals H2 and H6 are associated to an increased death hazard with respect to the average (see Figure~\ref{fig:Hosp3}).


The proposed model identified three distinct patient profiles within the cohort, allowing the estimation of cluster-specific covariate effects and frailty terms that influence the risk of hazard. This enabled the evaluation of how the impact of respiratory diseases and hospitals on the hazard varies between different patient profiles.

In particular, the impact of the hospital is most evident among generally healthy but vulnerable patients, specifically those in groups 2 and 3, where timely hospital care can significantly affect survival. This key insight comes from the model's ability to uncover heterogeneity in survival outcomes related to cluster-wise different respiratory conditions and hospital-specific factors among patient profiles. The model’s capability to capture such complex interactions is further demonstrated in a synthetic setting, as presented in the following section.

\begin{table}[htpb]
 \caption{Parameters estimated by the model with $G = 3$ clusters and a Lognormal baseline. The upper section presents the parameter estimates for the continuous covariates $\{\hat{\bm{\mu}}_g, \hat{\bm{\Sigma}}_g\}_{g = 1, \ldots, G}$, the categorical covariates $\{\hat{\bm{\pi}}_g\}_{g = 1, \ldots, G}$ and the number of patients in each cluster $\{\hat{n}_g\}_{g = 1, \ldots, G}$. The lower section provides the estimated fixed effects $\{\hat{\bm{\beta}}_g\}_{g = 1, \ldots, G}$, the estimated characteristic parameters of the baseline hazard Lognormal distribution $\{\hat{{\eta}}_g, {\hat{\nu}}_g\}_{g = 1, \ldots, G}$, and the estimated parameter of the $\text{Gamma}(1, \theta)$ distribution $\{\hat{\theta}_g\}_{g = 1, \ldots, G}$ employed to model the frailty term, along with their corresponding p-values. A Wald test was performed to assess whether the frailty parameter $\hat{\theta}_g$ differs significantly from zero in each cluster; the resulting approximate p-values are reported.}
    \centering
   \begin{tabular}{c  c  c  c  c  c  c}
    \hline
   \textbf{Parameter} &  \textbf{Cluster 1} &  \textbf{Cluster 2} &  \textbf{Cluster 3}   \rule[-0.4cm]{0cm}{1cm} \\
    \hline \hline
    $\hat{n}_g$ & $151$ &  $1071$ & $1864$  \rule[-0.4cm]{0cm}{1.1cm} \\
    $\hat{\bm{\mu}}_g$ & $(78.09; \ 31.79)$ &  $(81.79; \ 12.13)$ & $(81.38; \ 9.44)$  \rule[-0.4cm]{0cm}{1.1cm} \\
    $\hat{\bm{\Sigma}}_g$ & $\begin{bmatrix}
                62.69 & -2.26 \\
                -2.26 & 34.13 
                \end{bmatrix}$ & $\begin{bmatrix}
                64.28 & 0.68 \\
                0.68 & 39.56 
                \end{bmatrix}$ & $\begin{bmatrix}
                70.30 & 0.14 \\
                0.14 & 36.57
                \end{bmatrix}$   \rule[-1cm]{0cm}{2cm}\\
    $\hat{\bm{\pi}}_{\text{Gender}_g}$ & $(0.39; \ 0.61)$ &  $(0.44; \ 0.56)$ & $(0.45; \ 0.55)$  \rule[-0.4cm]{0cm}{-2cm} \\
    $\hat{\bm{\pi}}_{\text{COPD}_g}$ & $(0.76; \ 0.24)$ &  $(0; \ 1)$ & $(1; \ 0)$  \rule[-0.4cm]{0cm}{-2cm} \\
    $\hat{\bm{\pi}}_{\text{BRH}_g}$ & $(0.88; \ 0.12)$ &  $(0.56; \ 0.44)$ & $(0.97; \ 0.03)$  \rule[-0.4cm]{0cm}{-2cm} \\
    \hline
\end{tabular}
    \vspace{0.5cm} \\ 
    \renewcommand{\arraystretch}{1.3}
\begin{tabular}{cccccc c ccc ccc}
 \cline{3-4} \cline{6-7} \cline{9-10}
 \multicolumn{1}{c}{ }  && \multicolumn{2}{c}{\textbf{Cluster 1}} && \multicolumn{2}{c}{\textbf{Cluster 2}} && \multicolumn{2}{c}{\textbf{Cluster 3}}  \\
 \cline{1-1} \cline{3-4} \cline{6-7} \cline{9-10}
  \textbf{Parameter}  && \textbf{Estimate} & {$\textbf{Pr}(>|z|)$}  && \textbf{Estimate} & {$\textbf{Pr}(>|z|)$}    && \textbf{Estimate} & {$\textbf{Pr}(>|z|)$}  \\
  \cline{1-1} \cline{3-4} \cline{6-7} \cline{9-10}
  $\hat{\eta}_g$  &&  $-2.387$ &   && $-1.409$ &   && $-1.466$ &   \\
  $\hat{\nu}_g$  &&  $+1.754$ &   && $+2.118$ &   && $+2.131$ &   \\
   $\hat{\theta}_g$  &&  $+0.299$ & $0.19$  && $+0.139$ & $0.01$*  && $+0.188$ & $0.004$*   \\
 $\hat{\beta}_{PNA_g}$  &&  $-0.016$ & $0.94 $  && $+0.235$ & $0.02$*    && $+0.167$ & $0.08$*   \\
 $\hat{\beta}_{RF_g}$  &&  $+0.031$ & $0.91$  && $-0.043$ & $0.68$    && $+0.072$ & $0.54$   \\
\end{tabular} 
 \label{tab:app_parameters}
\end{table}

\begin{figure}[htpb]
    \centering
    \begin{subfigure}[b]{1\textwidth}
     \centering
        \includegraphics[width=0.9\textwidth]{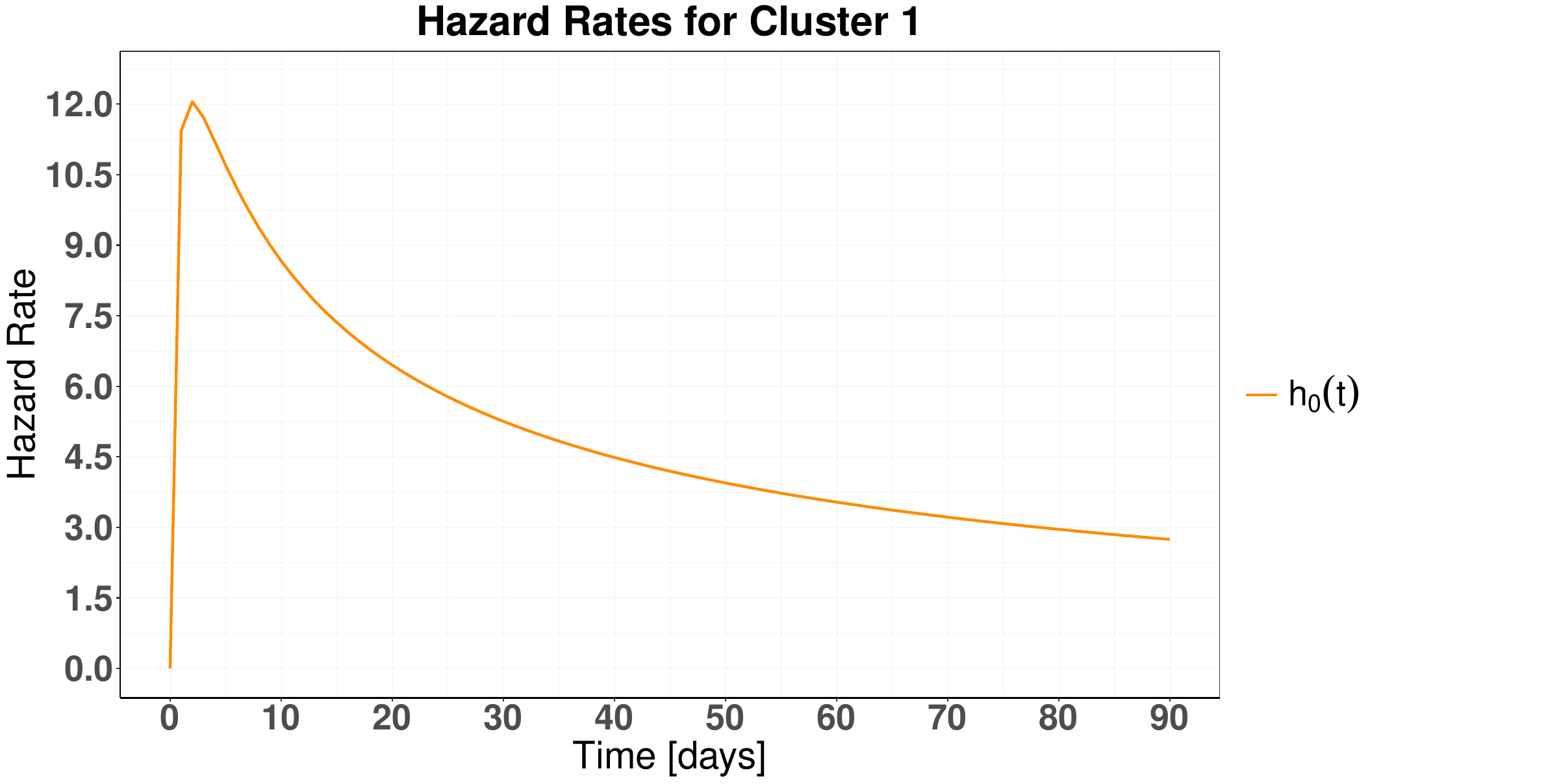}
        \label{fig: Haz1}
    \end{subfigure}
    \begin{subfigure}[b]{1\textwidth}
     \centering
        \includegraphics[width=0.9\textwidth]{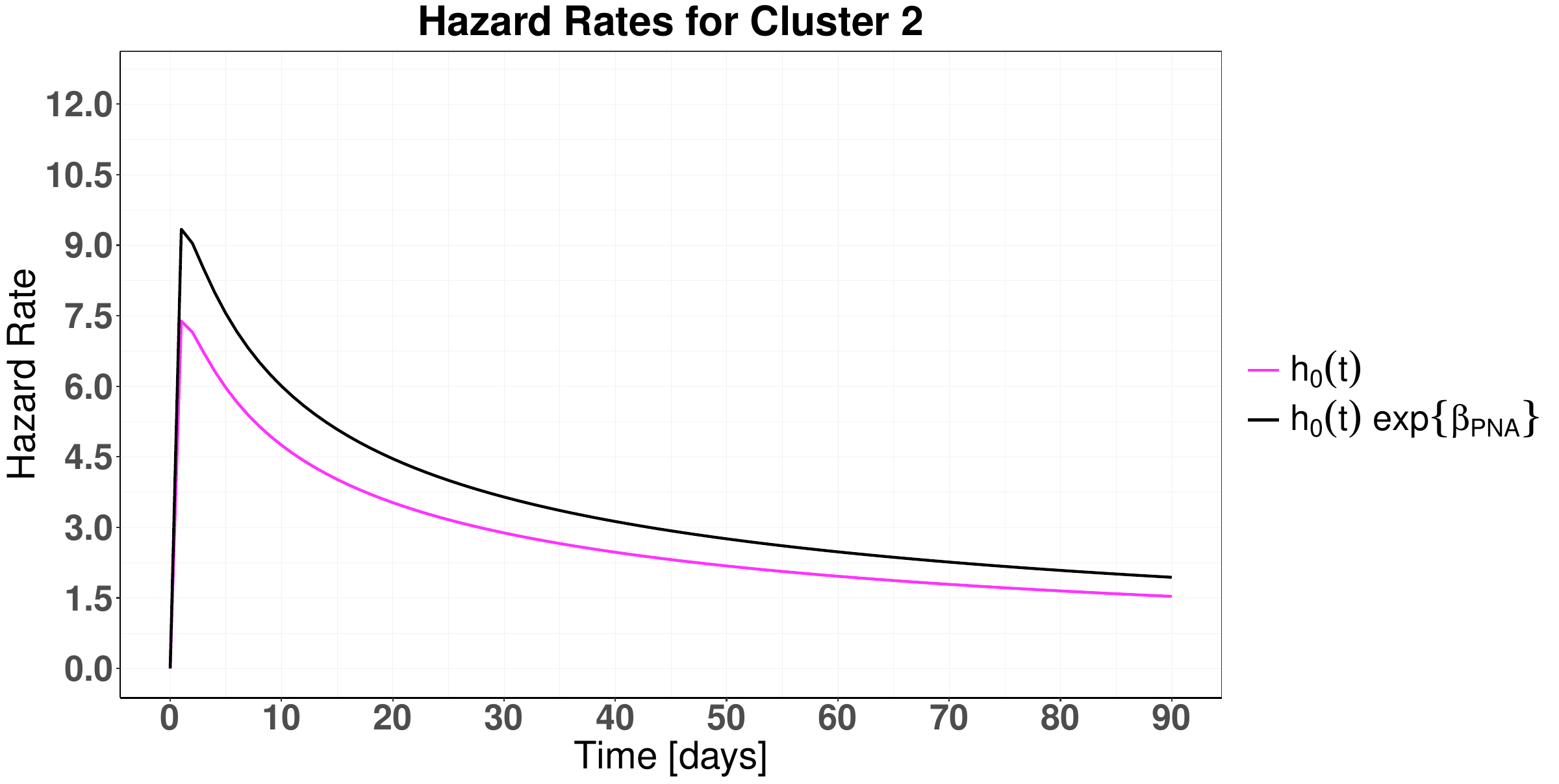} 
        \label{fig: Haz2}
    \end{subfigure}
    \begin{subfigure}[b]{1\textwidth}
     \centering
        \includegraphics[width=0.9\textwidth]{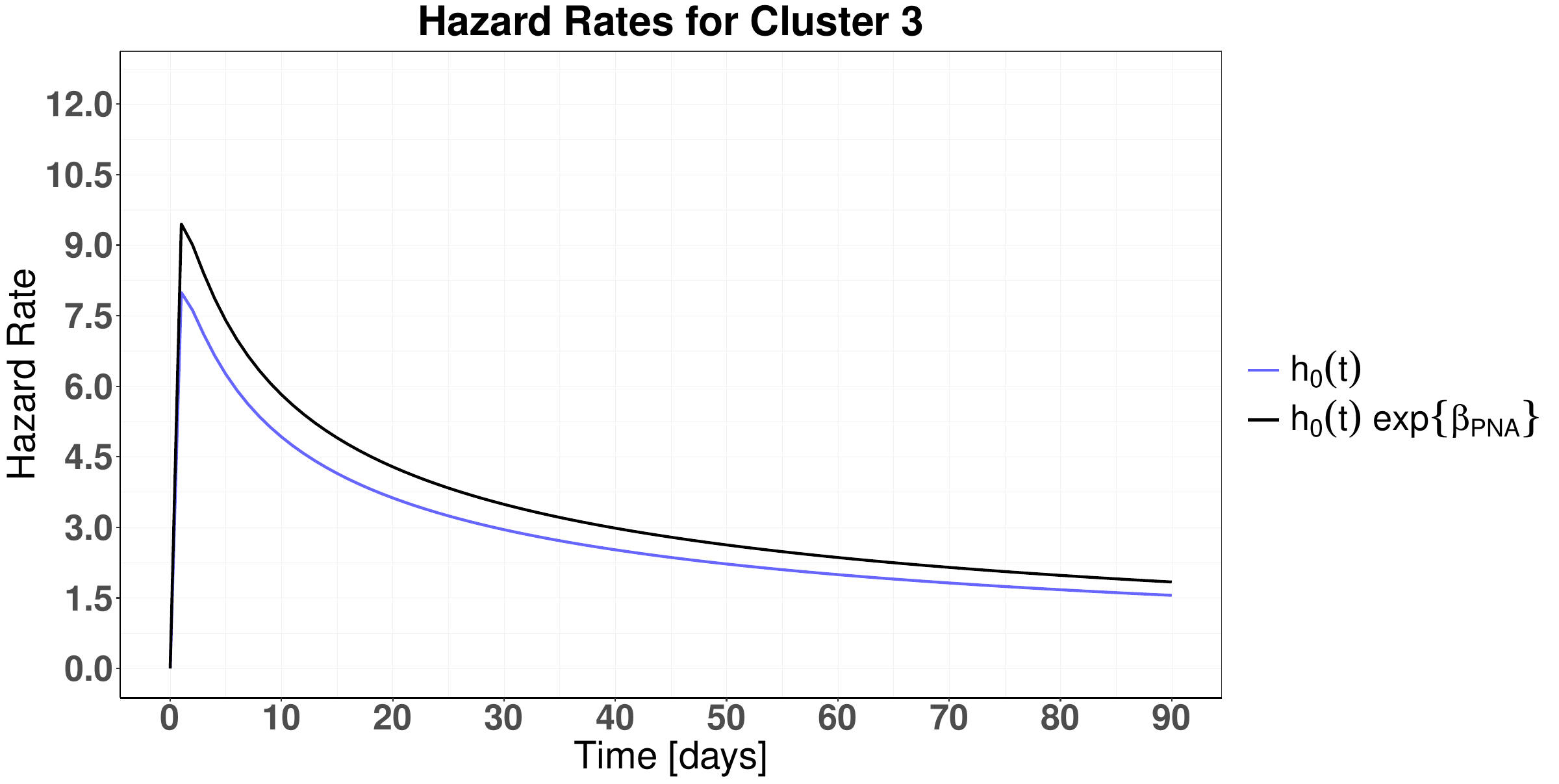}
        \label{fig: Haz3}
    \end{subfigure}
    \caption{Estimated baseline hazard in the three clusters and effect of the significant covariates (black curve).}
    \label{fig:app_hazard}
\end{figure}

\section{Simulation Study}
\label{sec:sim_study}
We hereafter present a comprehensive simulation study to show the performance of the method in a controlled setting. We consider a Data Generating Process (DGP) that incorporates various covariates, a known hierarchy,  and a latent grouping structure. Specifically, we consider the case of \(G = 3\) latent clusters and \(J = 10\) known groups. The total number of observations is set to \(N = 1500\), equally distributed among the 10 groups. Of the 150 observations from each group, 40, 50, and 60 belong to the first, second, and third cluster, respectively. 
The covariates include two continuous variables (\(p = 2\)) simulated from two independent Gaussian distributions with parameters \(\bm{\mu}_{g}\) and \(\bm{\sigma}_{g}\); one binary variable simulated using a Bernoulli distribution with parameter \(\boldsymbol{\pi}_{g1}\); and one categorical variable with three possible outcomes simulated using a Multinomial distribution with parameter \(\boldsymbol{\pi}_{g2}\), $g=1,\ldots,3$. 
A Weibull distribution is considered for the baseline hazard, with cluster-wise different shape \(\rho_g\) and scale \(\lambda_g\) parameters. 
The frailty distribution in each cluster is modeled using a Gamma density with a mean equal to 1 and unknown variance \(\theta_g\). To prevent frailty effects from dominating survival outcomes, we aim to keep the variance \(\theta_g\) relatively small. In the regression component, all covariates are included, resulting in a 5-dimensional coefficient vector $\boldsymbol{\beta}_g$ for each cluster $g$. Survival times are simulated by combining the baseline hazard with the effects of frailty and covariates. To generate the synthetic data, we utilize the \texttt{genfrail} function from the \texttt{frailtySurv} package \citep{Monaco2018}. The complete set of true parameter values used in the simulation is reported in Table~\ref{tab:sim3_params}. In the regression component, most of the coefficients for the first cluster are specified as risk factors, meaning that the majority of entries in $\boldsymbol{\beta}_1$ are positive. In contrast, the majority of coefficients in the third cluster are specified as protective factors, with most entries in $\boldsymbol{\beta}_3$ being negative. The second cluster exhibits a mixed profile, containing both risk and protective factors. High-risk clusters are assigned larger frailty variances, whereas low-risk clusters have smaller variances. The same shape parameter \(\rho_g = 3\), $\forall g=1,2,3$ is applied across clusters to simulate a quadratic growth in risk, with \(\lambda_g\) varying by cluster to create distinct risk profiles. The covariate distribution parameters vary between clusters, except for the variance of the continuous covariates \(\boldsymbol{\sigma}_g\), which is fixed at \(\{1, 1\}\) for all clusters.

We simulate $R = 100$ datasets following the DGP described above. For each dataset, we apply the proposed method by varying the number of clusters $G \in \{1, 2, 3, 4\}$.

\begin{table}[htpb]
\caption{Simulation parameters related to the survival part of the model and covariates distributions in each cluster.}
    \renewcommand{\arraystretch}{1.6}
\centering
\begin{tblr}{c c c c c c c}
\hline 
  \textbf{Cluster} & J & $n_g$ & $\boldsymbol{\beta}_g$ & $\theta_g$ & $\lambda_g$ & $\rho_g$ \\ 
 \hline \hline
$g = 1$ & 10 & 40 & $\{0.2, \, -0.1, \, 0.3, \, 0.5, \, 0.2 \}$ & 0.8 & 2  & 3 \\
\hline
$g = 2$ & 10 & 50 & $\{-0.2, \, -0.1, \, 0.2, \, -0.3, \, 0.15 \}$ & 0.6  & 0.7  & 3 \\
\hline
$g = 3$ & 10 & 60 & $\{-0.2, \, 0.2, \, -0.3, \, -0.3, \, -0.4 \}$ & 0.4  & 0.4  & 3 \\
\hline
\end{tblr}
    \vspace{0.5cm} \\ 
  \renewcommand{\arraystretch}{1.6}
\centering
\begin{tblr}{c c c c c c c}
\hline
  \textbf{Cluster} & $\boldsymbol{\mu}_g$ & $\boldsymbol{\sigma}_g$ & $\boldsymbol{\pi}_{g1}$ & $\boldsymbol{\pi}_{g2}$  \\ 
 \hline \hline
$g = 1$ & $\{1, \, -3 \}$ & $\{1, \, 1 \}$ & $\{0.4, \, 0.6 \}$ & $\{0.3, \, 0.5, \, 0.2 \}$ \\
\hline
$g = 2$ & $\{3, \, 1 \}$ & $\{1, \, 1 \}$ & $\{0.8, \, 0.2 \}$ & $\{0.6, \, 0.1, \, 0.3 \}$   \\
\hline
$g = 3$ & $\{5, \, 3 \}$ & $\{1, \, 1 \}$ & $\{0.2, \, 0.8 \}$ & $\{0.1, \, 0.3, \, 0.6 \}$   \\
\hline
\end{tblr}
\label{tab:sim3_params}
\end{table}

\subsection{Results}
\label{subsec: results g3 sim}
\begin{figure}[htpb]
\centerline{%
\includegraphics[width=\textwidth]{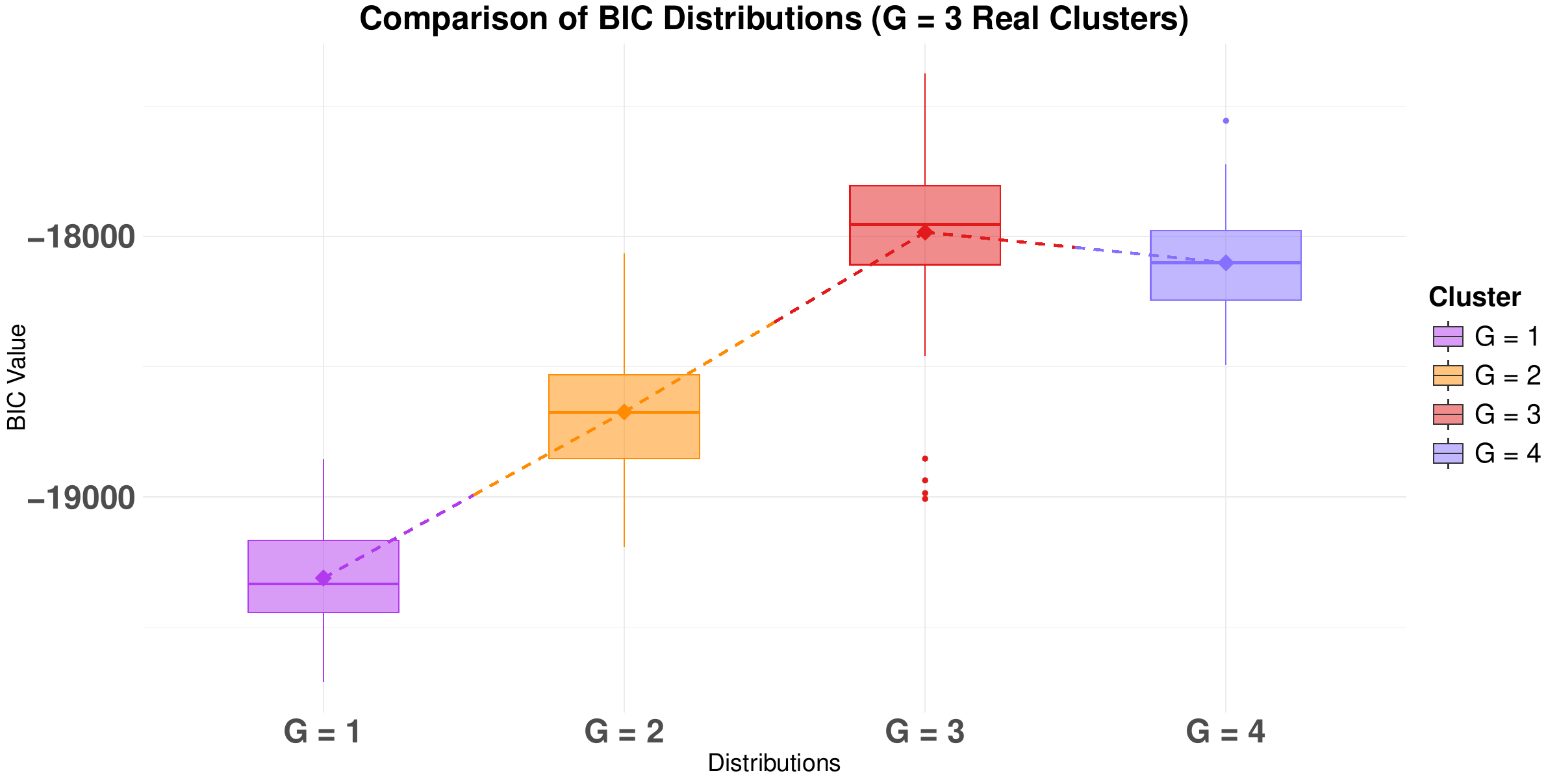}}
\caption{Comparison of BIC distributions for \( G \in \{1, 2, 3, 4\} \). Each boxplot represents the BIC distribution for a given \( G \) across the 100 repetitions of the simulated experiment.}
\label{fig: fir g3 BIC}
\end{figure}

\begin{figure}[H]
\centerline{%
\includegraphics[width=\textwidth]{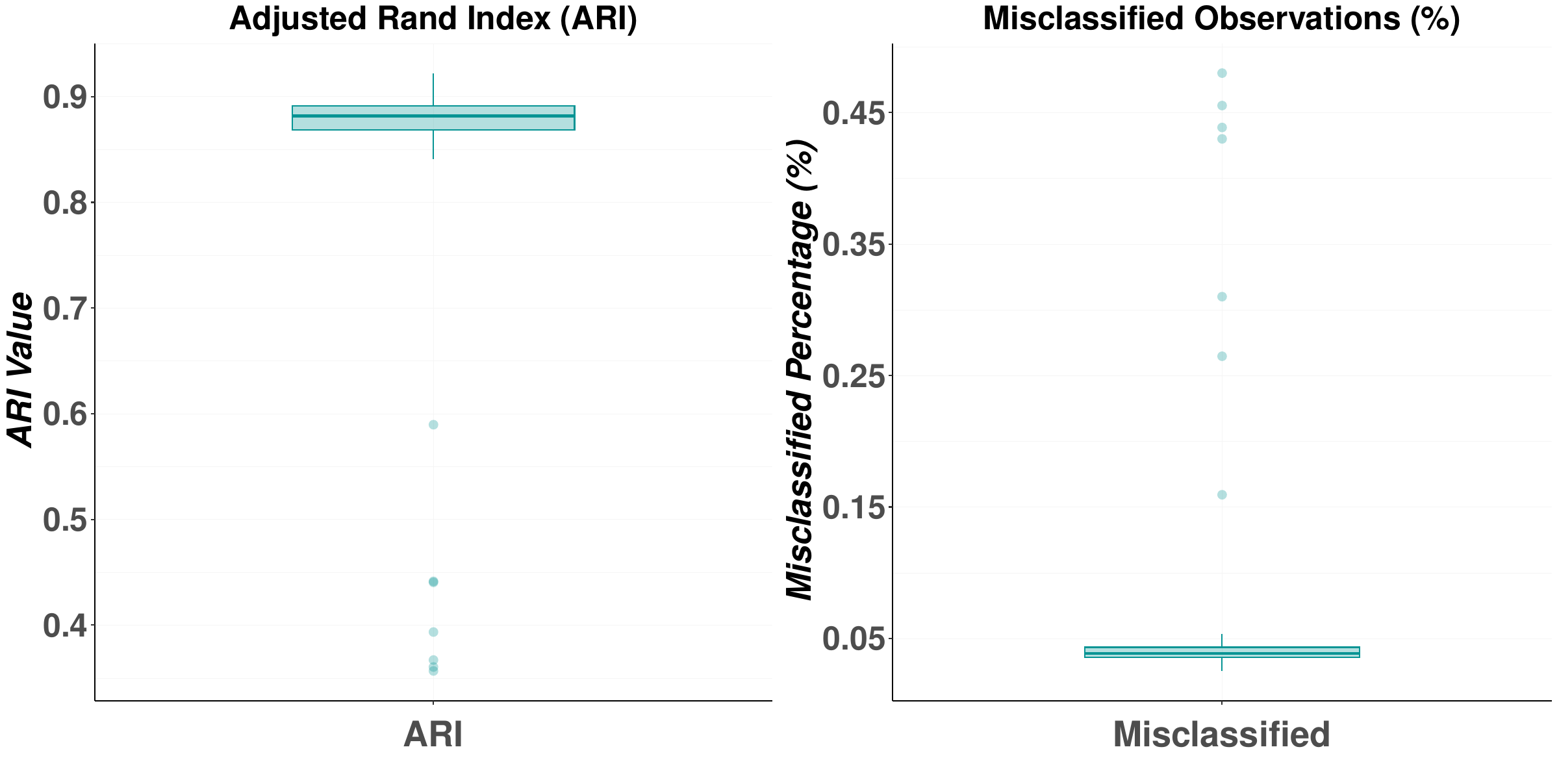}}
\caption{ARI and percentage of missclassified observations across the 100 simulations.}
\label{fig:clust_acc}
\end{figure}

For each repetition of the simulated experiment, we use the BIC to assess which model is preferred by varying the number of components $G \in \{1, 2, 3, 4\}$. 
The empirical BIC distributions shown in Figure~\ref{fig: fir g3 BIC} indicate that the model with G = 3 clusters is preferred. 
Considering $G=3$ as the best solution, to evaluate the ability of the model to recover latent clusters, we examine the misclassification rates and the Adjusted Rand Index (ARI)  \citep{Hubert1985} across the 100 runs (Figure ~\ref{fig:clust_acc}). In 93 of 100 runs, the model demonstrates effective recovery of latent clusters, achieving an average misclassification rate of $0.039$ ($SD=0.005$) and an ARI of $0.882$ ($SD=0.016$). 

\begin{figure}[htpb]
    \centering
    \begin{subfigure}[b]{0.8\textwidth}
        \includegraphics[width=\textwidth]{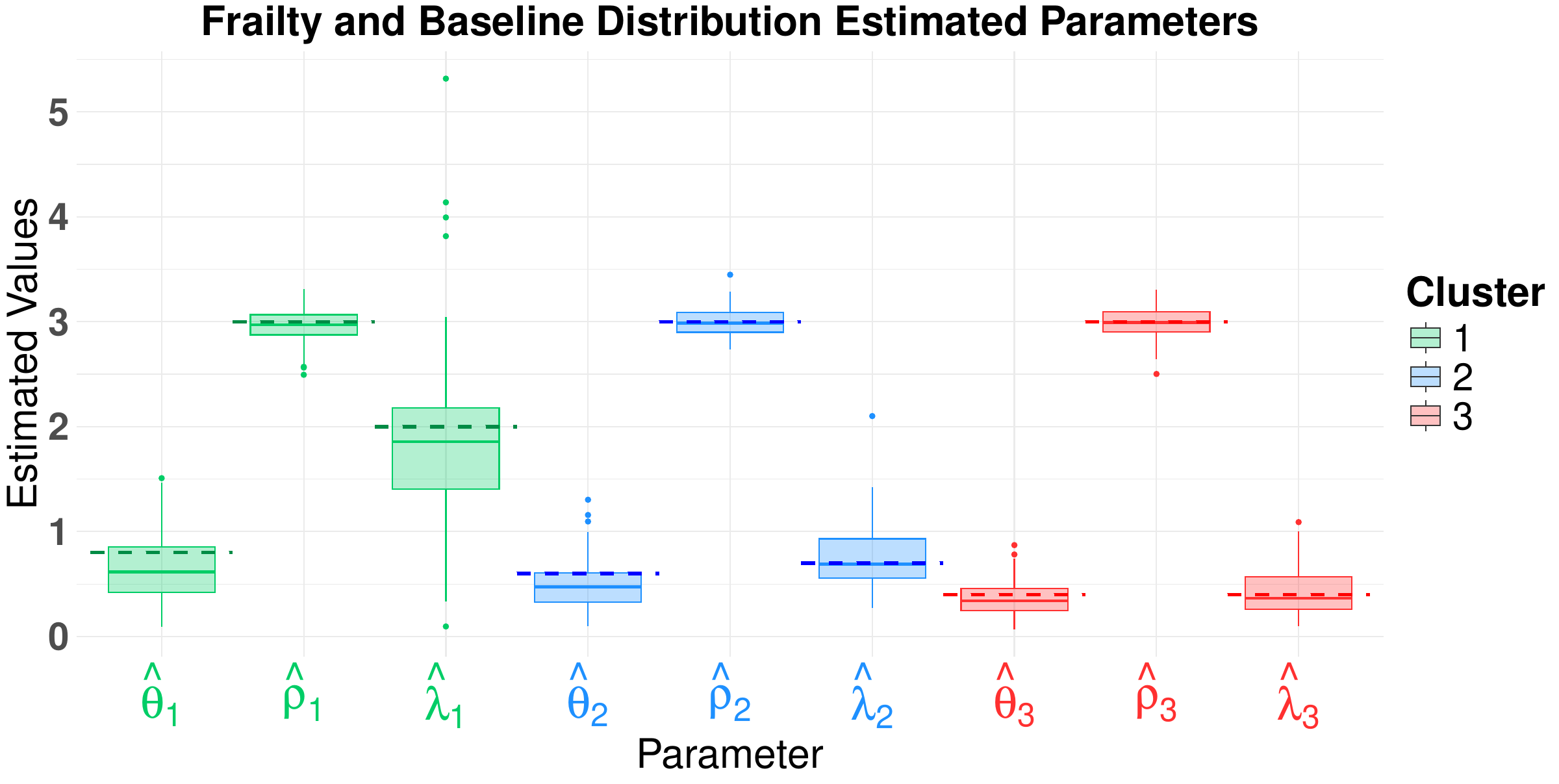}
        \caption{Parameters related to the frailty distribution ($\{\theta_g \}_{g = 1}^G$) and the baseline distribution ($\{ \rho_g, \lambda_g \}_{g = 1}^G$).}
        \label{fig:fir g3 distr}
    \end{subfigure}
    \hfill
    \begin{subfigure}[b]{0.8\textwidth}
        \includegraphics[width=\textwidth]{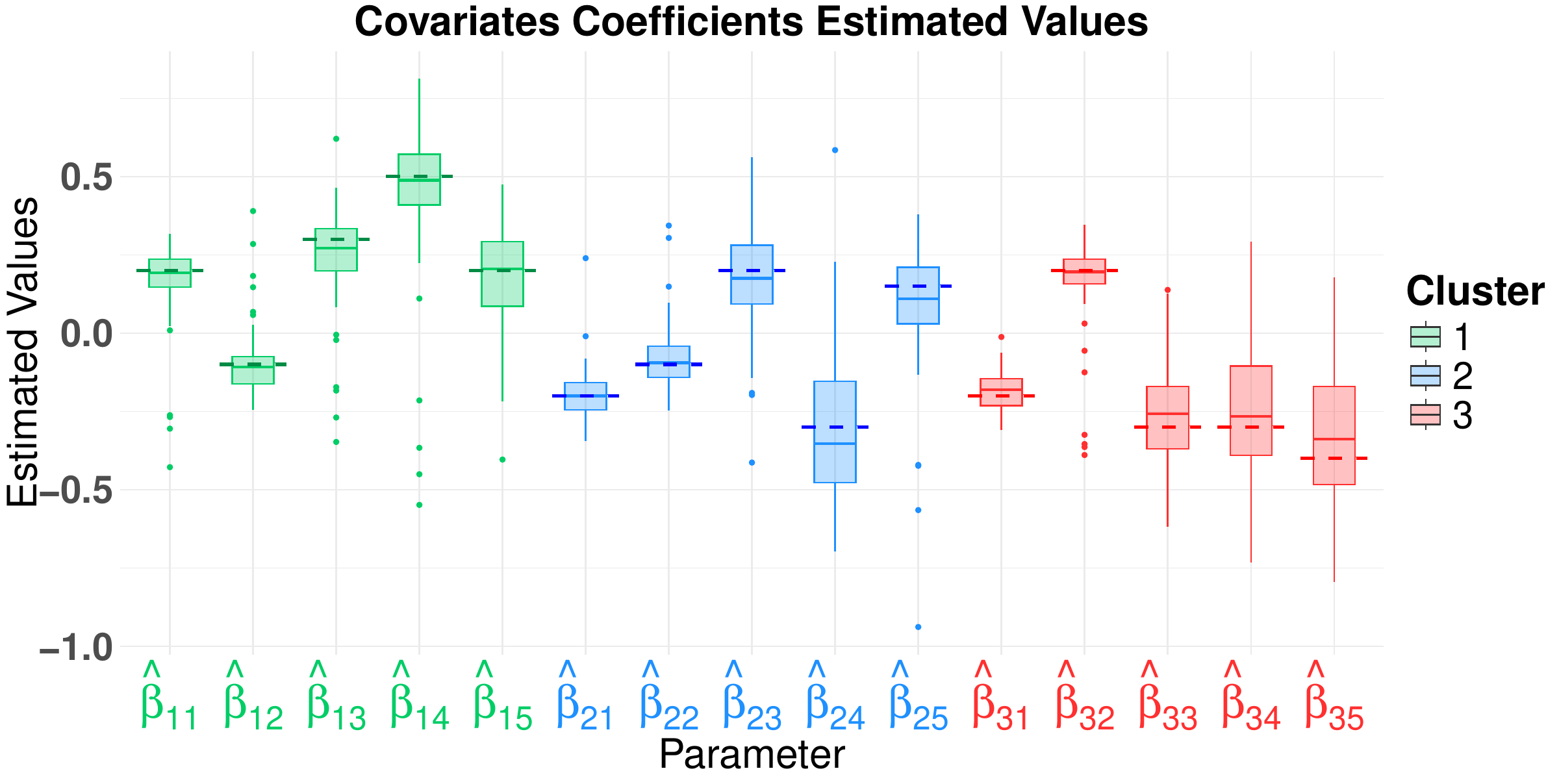} 
        \caption{Parameters related to the covariates coefficients ($\{\beta_{gk} \}_{g = 1, \, \hdots, \, G , \  k = 1, \, \hdots, \, m}$).}
        \label{fig:fir g3 beta}
    \end{subfigure}
    \hfill
    \begin{subfigure}[b]{0.8\textwidth}
    \centering
        \includegraphics[width=\textwidth]{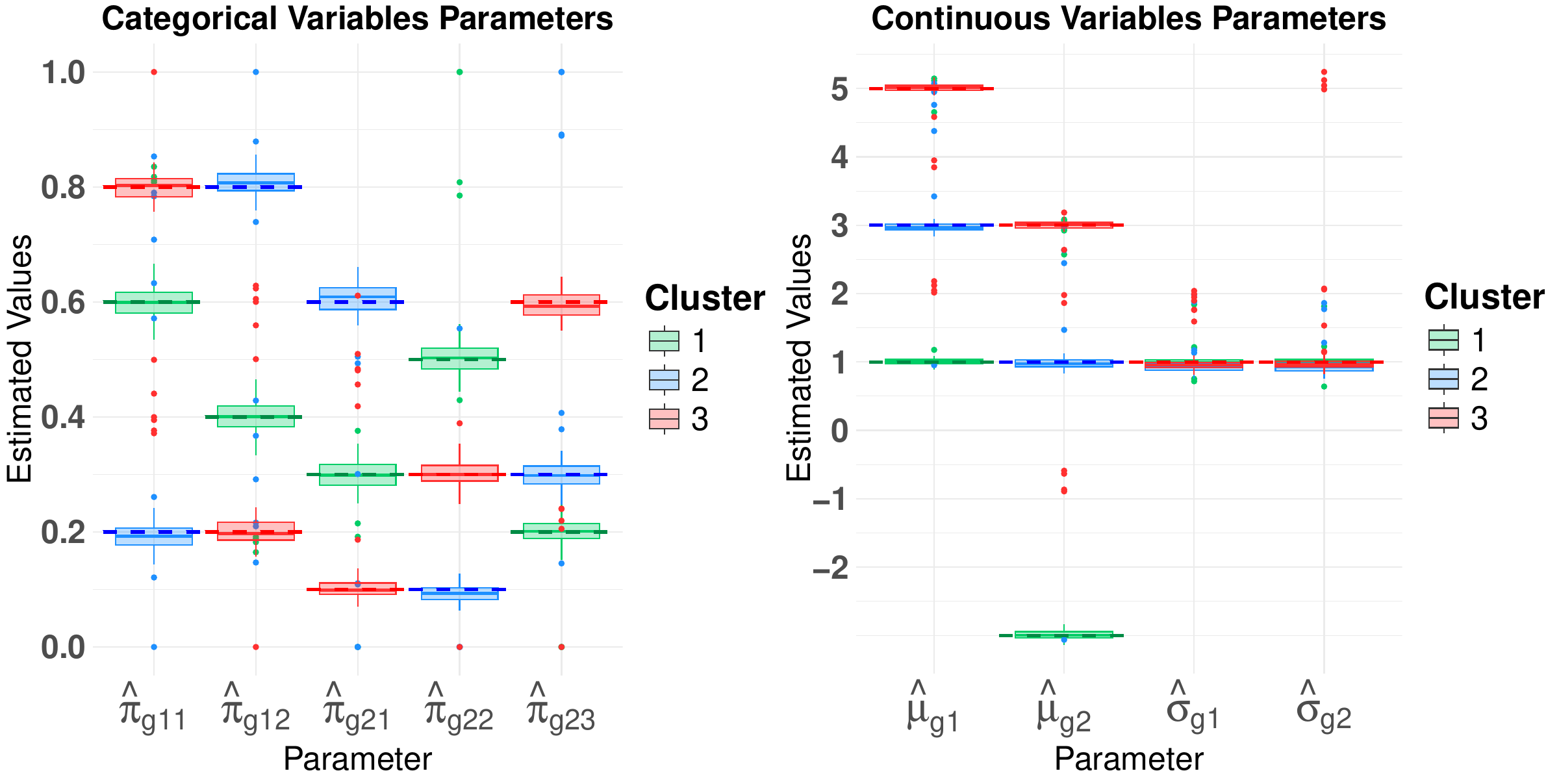}
        \caption{Parameters related to the covariates distributions ($\{\mu_{gh}, \sigma_{gh}  \}_{g = 1, \, \hdots, \, G , \ h = 1, \, \hdots, \, p}$ for continuous covariates and $\{\boldsymbol{\pi}_{grl}  \}_{g = 1, \, \hdots, \, G , \ r = 1, \, \hdots, \, q, \  l = 1, \, \hdots, \, k_r}$ for categorical covariates).}
        \label{fig:fir g3 cov}
    \end{subfigure}
    \caption{Comparison of the distributions of estimated parameters across the R=100 repetitions with the true values of DGP. Dashed lines represent the true values.}
    \label{fig:firg3}
\end{figure}

In Figure ~\ref{fig:firg3}, we compare the distributions of the estimated parameters of the simulation experiments with the true values for the case $G=3$. For what concerns the survival component (Figure ~\ref{fig:fir g3 distr}), the parameters specific to each of the three clusters are estimated with high precision.
However, the frailty variance is slightly underestimated in all three clusters.  Regarding the covariates coefficients (Figure ~\ref{fig:fir g3 beta}), their estimates within each cluster are highly precise, effectively capturing the numerical values of risk and protective factors. Lastly, the parameters governing the random covariates are also well-estimated in most simulations (Figure ~\ref{fig:fir g3 cov}). 

Other simulation studies, such as those that involve different numbers of latent clusters, alternative specifications of the baseline hazard function, or larger sample sizes, are certainly possible. In this work, we focus on a standard scenario that incorporates all the key components addressable by the proposed approach. The results demonstrate the strong performance of the estimation method, showing a high precision in recovering both the model parameters and the latent clusters. These findings provide a solid foundation while leaving ample room for future investigations in more complex or varied settings.

\section{Discussion}
\label{sec:discussion}
This paper introduced a novel methodology that extends the CWM framework to accommodate hierarchical time-to-event outcomes. The core innovation lied in the ability of the methodology to address both the clustering of multilevel data and the inherent heterogeneity in survival analysis concurrently. 
The proposed methodology identified clusters with distinct survival patterns by estimating cluster-specific coefficients that influence hazard risks. This enabled the assessment of how various factors impact the hazard within each cluster. In addition, the method provided cluster-specific frailty estimates, capturing unobserved heterogeneity in survival outcomes and accounting for deviations from the baseline risk due to unmeasured factors. Its fully parametric structure allowed for flexible specification of both the baseline hazard and frailty distributions.  Two EM-based algorithms, tailored for right-censored data, were devised for parameter estimation: one incorporating a classification step (CEM algorithm) and the other relying on a stochastic step (SEM algorithm).

This work was motivated by the analysis of a real-world administrative dataset from the Lombardy region in Italy, focusing on patients with heart failure (HF) who were hospitalized with COVID-19. Our proposed methodology enabled meaningful insights into the interaction between HF and COVID-19 during the pandemic. The analysis identified three distinct patient profiles, each exhibiting unique survival patterns. Furthermore, the approach allowed for the evaluation of respiratory conditions and hospital-level effects on individual patient profiles through cluster-specific estimates of covariate coefficients and frailty terms, respectively. These findings demonstrate the potential of the methodology to generate actionable insights for healthcare planning. By uncovering survival trends and key influencing factors, the method highlights opportunities to optimize the treatment of HF patients at the territorial level, ultimately with the aim of reducing adverse outcomes and improving the efficiency of the healthcare system.

A limitation of the proposed approach is the assumption that each patient consistently belongs to a single latent profile. While this may be reasonable in cross-sectional settings, it can be overly restrictive in longitudinal contexts, where patients’ characteristics and risk profiles may change over time. Relaxing this constraint represents an important avenue for future methodological development.

Further future extensions of the proposed methodology may involve integrating the Ising model into the marginal distributions of the covariates to capture dependencies between binary variables, such as comorbidities or treatment indicators, commonly encountered in clinical and administrative datasets, along the lines of \cite{Caldera2025}. 
The proposed framework could also be adapted to accommodate high-dimensional data through the integration of sparsity-inducing estimation techniques. This extension would enable effective variable selection by identifying the most important covariates within each cluster, while also mitigating overfitting and improving generalizability. Developing penalized likelihood formulations specifically adapted to the cluster-weighted model of survival data would be particularly valuable for large-scale applications, such as those involving electronic health records or genomics, where the number of predictors may largely exceed the sample size. Several such approaches are currently under investigation and will be the focus of future work.

\section*{Acknowledgments}
This work is part of the ENHANCE-HEART project: Efficacy evaluatioN of the therapeutic-care patHways, of
the heAlthcare providers effects aNd of the risk stratifiCation in patiEnts suffering from HEART failure. The authors thank the ‘Unità Organizzativa Osservatorio Epidemiologico Regionale’ and ARIA S.p.A for providing data and technological support. The authors gratefully acknowledge the support from the Department of Mathematics of Politecnico di Milano, which facilitated this research as part of the department's activities of ``Dipartimento di Eccellenza 2023-2027''. Chiara Masci acknowledges financial support from the Italian
Ministry of University and Research (MUR) under the Department of Excellence 2023-2027 grant
agreement “Centre of Excellence in Economics and Data Science” (CEEDS). \vspace*{-8pt}

\newpage
\huge{\textbf{Appendix}}
\normalsize
\section*{Appendix A: Further details pertaining the model definition}

\subsection*{Laplace Transform and Derivatives of the Frailty Distribution }

For a positive random variable $M$ its Laplace transform is defined as:
\begin{align}
\label{eq: lapl trans}
    \mathcal{L}(s) = \mathbbm{E}[e^{- sM}] = \int_{0}^{\infty} e^{-sm}f_M(m) \ dm,
\end{align}
where $f_M(\cdot)$ is the probability density function of M, and $s$ is a non-negative real number.
The first derivative of $\mathcal{L}(s)$ with respect to $s$ can be computed as follows:
\begin{align}
\label{eq: der lapl trans 1}
    \frac{d}{ds} \mathcal{L}(s) &= \frac{d}{ds} \int_{0}^{\infty} e^{-sm}f_M(m) \ dm = \text{Leibniz integral rule} \notag \\
   &= \int_{0}^{\infty} f_M(m) \frac{d}{ds}e^{-sm} \ dm = - \int_{0}^{\infty} m f_M(m) e^{-sm}\ dm.
\end{align}
Differentiating Equation~\eqref{eq: der lapl trans 1} again, we obtain the second derivative of $\mathcal{L}(s)$ with respect to $s$:
\begin{align}
\label{eq: der lapl trans 2}
    \frac{d^2}{ds^2} \mathcal{L}(s) = \int_{0}^{\infty} m^2 f_M(m) e^{-sm}\ dm.
\end{align}
Therefore, the general expression of the $q$-th derivative of the Laplace transform $\mathcal{L}(s)$ is given by:
\begin{align}
\label{eq: der lapl trans 3}
   \frac{d^q}{ds^q} \mathcal{L}(s) =  \mathcal{L}^{(q)}(s) =(-1)^q \int_{0}^{\infty} m^q f_M(m) e^{-sm}\ dm.
\end{align}
In our case, the random variable $M$ represents the frailty distribution which is parameterized by its variance $\theta$. Therefore, we denote the $q$-th derivative of the Laplace transform as dependent on $\theta$:
\begin{align}
\label{eq: der lapl trans 4}
   \frac{d^q}{ds^q} \mathcal{L}(s; \theta) =\mathcal{L}^{(q)}(s; \theta) =(-1)^q \int_{0}^{\infty} m^q f_M(m; \theta) e^{-sm}\ dm.
\end{align}

\subsection*{Derivation of the classification log-likelihood}
Starting from the classification likelihood defined in Equation (6) of the main paper the associated classification log-likelihood is derived as follows:
\begin{align}
     \begin{split}
     \label{eq:first_piece}
\ell\left(\boldsymbol{\psi}\right) =\log{L(\bm{\psi})}=&\sum_{g=1}^G\left\{\sum_{j=1}^J \log{\left( \int_{0}^{+\infty} \prod_{i \in R_{jg}}h(y_{ij} | m_{jg},\boldsymbol{x}_{ij}; \boldsymbol{\gamma}_g, \boldsymbol{\beta}_g)^{\delta_{ij}} S(y_{ij} | m_{jg},\boldsymbol{x}_{ij}; \boldsymbol{\gamma}_g, \boldsymbol{\beta}_g) f_M(m_{jg};\theta_g) dm_{jg}\right)} \right.+\\
&\left.  \sum_{j=1}^J\sum_{i \in R_{jg}} \left(  \log{\tau_g} + \log{\phi(\mathbf{u}_{ij}; \boldsymbol{\mu}_g,\boldsymbol{\Sigma}_g)}+ \log{\xi(\boldsymbol{v}_{ij}; \boldsymbol{\pi}_g)} \right)       \right\}.
\end{split}
\end{align}
where we have explicitly reported the expression of $L_{jg}^S(\boldsymbol{\gamma}_g, \boldsymbol{\beta}_g,\theta_g)$ as per Equation (5) of the main paper.
The second line of Equation \eqref{eq:first_piece} corresponds to the CWM contribution for the mixing proportion and the random covariates, and it directly aligns with the third row of Equation (7) in the main paper. Let us focus on the first line of Equation \eqref{eq:first_piece} corresponding to the contribution of the parametric frailty model for the observations in group $j$ assigned to the $g$-th component:

\begin{align}
     \begin{split}
     \label{eq:PFM_piece1}
   &\log{\left( \int_{0}^{+\infty} \prod_{i \in R_{jg}}h(y_{ij} | m_{jg},\boldsymbol{x}_{ij}; \boldsymbol{\gamma}_g, \boldsymbol{\beta}_g)^{\delta_{ij}} S(y_{ij} | m_{jg},\boldsymbol{x}_{ij}; \boldsymbol{\gamma}_g, \boldsymbol{\beta}_g) f_M(m_{jg};\theta_g) dm_{jg}\right)}=\text{Eq(2) and (3) of the main paper}  \\
   =&\log{\left( \int_{0}^{+\infty} \prod_{i \in R_{jg}}\left(h_0(y_{ij}; \boldsymbol{\gamma}_g) m_{jg} \exp\{\boldsymbol{x}_{ij}^T \, \boldsymbol{\beta}_g \}\right)^{\delta_{ij}} \exp\{-m_{jg} \exp\{\boldsymbol{x}_{ij}^T \, \boldsymbol{\beta}_g \} H_0(y_{ij}; \boldsymbol{\gamma}_g)\} f_M(m_{jg};\theta_g) dm_{jg}\right)}=\\
   =&\log{ \left(\prod_{i \in R_{jg}} \left( h_0(y_{ij}; \boldsymbol{\gamma}_g) \exp\{\boldsymbol{x}_{ij}^T \, \boldsymbol{\beta}_g \}\right)^{\delta_{ij}}\int_{0}^{+\infty}  m_{jg}^{\sum_{i \in R_{jg}}\delta_{ij}}  \exp\left\{-m_{jg} \sum_{i \in R_{jg}}\exp\{\boldsymbol{x}_{ij}^T \, \boldsymbol{\beta}_g \} H_0(y_{ij}; \boldsymbol{\gamma}_g)\right\} f_M(m_{jg};\theta_g) dm_{jg}\right)}.
          \end{split}
 \end{align}
Denoting with $s=\sum_{i \in R_{jg}}\exp\{\boldsymbol{x}_{ij}^T \, \boldsymbol{\beta}_g \} H_0(y_{ij}; \boldsymbol{\gamma}_g)$ and $d_{jg}=\sum_{i \in R_{jg}}\delta_{ij}$ the expression in Equation \eqref{eq:PFM_piece1} can be rewritten as:
\begin{align}
     \begin{split}
     \label{eq:PFM_piece2}
     &\log{ \left(\prod_{i \in R_{jg}} \left( h_0(y_{ij}; \boldsymbol{\gamma}_g) \exp\{\boldsymbol{x}_{ij}^T \, \boldsymbol{\beta}_g \}\right)^{\delta_{ij}}\int_{0}^{+\infty}  m_{jg}^{d_{jg}}  \exp\left\{-m_{jg} s\right\} f_M(m_{jg};\theta_g) dm_{jg}\right)}=\text{Eq \eqref{eq: der lapl trans 4}}\\
     =&\sum_{i\in R_{jg}}\delta_{ij}\left( \log{h_0(y_{ij}; \boldsymbol{\gamma}_g)} + \boldsymbol{x}_{ij}^T \, \boldsymbol{\beta}_g\right)+ \log{\left( (-1)^{d_{jg}} \mathcal{L}^{(d_{jg})}(s; \theta_g)\right)}=\\
     =&\sum_{i\in R_{jg}}\delta_{ij}\left( \log{h_0(y_{ij}; \boldsymbol{\gamma}_g)} + \boldsymbol{x}_{ij}^T \, \boldsymbol{\beta}_g\right)+ \log{\left( (-1)^{d_{jg}} \mathcal{L}^{(d_{jg})}\left(\sum_{i \in R_{jg}}\exp\{\boldsymbol{x}_{ij}^T \, \boldsymbol{\beta}_g \} H_0(y_{ij}; \boldsymbol{\gamma}_g); \theta_g\right)\right)}
           \end{split}
\end{align}
By substituting the expression obtained from Equation \eqref{eq:PFM_piece2} into Equation \eqref{eq:first_piece}, we derive the classification log-likelihood presented in Equation (7) of the main paper.

\section*{Appendix B: Further details pertaining the application to Lombardy region data}
In this appendix, we include additional information related to the application of the proposed model for profiling COVID-19 Heart Failure patients (Section 4 of the main paper).

\subsection*{Description of clinical variables involved in the analysis}
In the following, we provide a brief description of the respiratory diseases considered in the analysis and of the Multi-source Comorbidity Score (MCS). 

\noindent
The Chronic obstructive pulmonary disease (COPD) entails a persistent constriction or blockage of the air passages resulting in an ongoing reduction in the airflow rate during exhalation. Pneumonia (PNA) refers to the abrupt inflammation of the lungs brought about by an infection. Respiratory failure (RF) manifests when the oxygen level in the blood becomes critically low or when there is a dangerous elevation in the blood's carbon dioxide level. Bronchitis (BRH) entails an inflammation of the primary air passages of the lungs, known as the bronchi, typically triggered by infection, leading to irritation and inflammation.

\noindent
For what concerns the computation of the MCS, the scores associated to single diseases are presented in Web Table~\ref{table MCS 1}. The overall score for each patient is computed by summing the scores associated with all the diseases they have been diagnosed with. The diseases attributed to each patient are identified by analyzing their medical history pertaining the previous 5 years, a process facilitated through the utilization of ICD-9-CM codes. These codes serve as a means to extract relevant medical information and establish connections to specific diseases. 
\subsection*{ICD-9-CM codes for respiratory diseases}
\begin{itemize}
 \item \textbf{Chronic obstructive pulmonary disease (COPD)}: 
 \texttt{491-496, 491.2, 492.0, 492.8, 494.0-494.1}
    \item \textbf{Pneumonia (PNA):} \texttt{480-486, 507, 011.6, 052.1, 055.1, 073.0, 130.4, 480.0-480.3 \\, 480.8-480.9, 482.1-482.4, 482.8-482.9, 483.1, 483.8, 484.3, 484.5, 487.0, 506.0, 507.0, 507.8, 517.1, 770.0, 00322, 011.61-011.66, 115.05, 115.15, 115.95, \\ 482.30-482.32, 482.40-482.41, 482.49, 482.81, 482.89, V12.61}
    \item \textbf{Respiratory failure (RF):} \texttt{518.81, 518.83-518.84}
    \item \textbf{Bronchitis (BRH):} \texttt{466, 490-491,  466.0, 491.0-491.2, 491.8-491.9, 491.20-491.22}
\end{itemize}

\begin{table}[htpb]
\caption{The score associated to each disease to define the Modified Multisource-Comorbidity Score (MCS).}
\begin{minipage}{0.5\textwidth}
\centering
\begin{tabular}{ l c}
\hline
\rowcolor{gray!20}
\textbf{Comorbidity} & \textbf{Score} \\
\hline \hline
Metastatic cancer & 18  \\
\hline
Alcohol abuse & 11 \\
\hline
Non-metastatic cancer & 10 \\
  \hline
    Tuberculosis & 10    \\
    \hline
    Psychosis & 8     \\
    \hline
    Liver diseases & 8  \\
    \hline
    Drugs for anxiety & 6    \\
    \hline
    Weight loss & 6    \\
    \hline
    Dementia & 6    \\
    \hline
    Drugs for malignancies & 5    \\
    \hline
    Parkinson's disease & 5   \\
    \hline
    Lymphoma & 5   \\
    \hline
    Paralysis & 5    \\
    \hline
    Coagulopathy & 5   \\
    \hline
    Fluid disorders & 4    \\
    \hline
    Kidney diseases & 4    \\
    \hline
\end{tabular}
\label{table MCS 1}
\end{minipage}%
\begin{minipage}{0.5\textwidth}
\centering
\begin{tabular}{ l c}
\hline
\rowcolor{gray!20}
\textbf{Comorbidity} & \textbf{Score}  \\
\hline \hline
 Kidney dialysis & 4    \\
    \hline
    Heart failure & 4    \\
    \hline
    Other neurological disorders & 3    \\
    \hline
    Rheumatic diseases & 3    \\
    \hline
    Brain diseases & 3   \\
    \hline
     Anemia & 3    \\
    \hline
    Diabetes & 2    \\
    \hline
    Gout & 2    \\
    \hline
    Epilepsy & 2    \\
    \hline
    Ulcer diseases & 2    \\
    \hline
Myocardial infarction & 1 \\
\hline
Drugs for coronary & 1 \\
\hline
Valvular diseases & 1 \\
\hline
Arrhythmia & 1 \\
\hline
Obesity & 1 \\
\hline
Hypothyroidism & 1 \\
\hline
\end{tabular}
\label{table MCS 2}
\end{minipage}
\end{table}

\section*{Appendix C: Computation of survival function confidence intervals}
In order to compute confidence intervals for the survival function in Section 4.2 of the main paper, we rely on the delta method \citep{ver2012invented}, which is a technique for approximating the variance of a non-linear function of random variables.

In our case, the survival function depends on the parameters \( \eta_g \) and \( \nu_g \) for each cluster $g \in \{1,2,3\}$. To apply the delta method, we first compute the partial derivatives of the survival function with respect to these parameters:
\[
\frac{\partial S(t)}{\partial \eta_g} = -\frac{1}{\nu_g} \phi\left( \frac{\log(t) - \eta_g}{\nu_g} \right); \quad \frac{\partial S(t)}{\partial \nu_g} = -\frac{\log(t) - \eta_g}{\nu_g^2} \phi\left( \frac{\log(t) - \eta_g}{\nu_g} \right)
\]
Next, the variance of the survival function at a specific time \( t \) is calculated using the following formula:
\[
\text{Var}(S(t)) = \left( \frac{\partial S(t)}{\partial \eta_g} \right)^2 \text{Var}(\eta_g) + \left( \frac{\partial S(t)}{\partial \nu_g} \right)^2 \text{Var}(\nu_g)
\]
where, \( \text{Var}(\eta_g) \) and \( \text{Var}(\nu_g) \) are the squared standard errors of the fitted parameters \( \eta_g \) and \( \nu_g \), respectively. This formula accounts for the uncertainty in the survival function resulting from the variability in the estimates of \( \eta_g \) and \( \nu_g \). The standard error of the survival function is then the square root of the variance:
\[
\text{SE}(S(t)) = \sqrt{\text{Var}(S(t))}.
\]
Finally, we construct the 95\% confidence intervals for the survival function at time \( t \) using the standard normal quantile \( Z = 1.96 \) for a two-sided confidence level:
\[
\text{CI}_{\text{95\%}} = S(t) \pm Z \cdot \text{SE}(S(t)).
\]
This approach provides a method to quantify the uncertainty around our estimate of the survival function at a given time, based on the estimated parameters and their associated standard errors.

\bibliographystyle{apalike}

\end{document}